\documentstyle[12pt]{article}
\newcommand{\beq}{\begin{equation}}
\newcommand{\eeq}{\end{equation}}
\newcommand{\ber}{\begin{eqnarray}}
\newcommand{\eer}{\end{eqnarray}}

\newcommand{\eeql}[1]{\label{#1}\eeq} 
\newcommand{\eerl}[1]{\label{#1}\eer} 

\newcommand{\teq}{\hskip-.1in&=&\hskip-.1in}
\newcommand{\tab}{\hskip-.1in&&\hskip-.1in}

\newcommand{\vev}[1]{{\left< {#1} \right>}}
\newcommand{\bra}[1]{{\left< {#1} \right|}}
\newcommand{\ket}[1]{{\left| {#1} \right>}}

\newcommand{\Tr}{{\rm Tr}}
\newcommand{\cN}{{\cal N}}
\newcommand{\cC}{{\cal C}}
\newcommand{\cD}{{\cal D}}
\newcommand{\cL}{{\cal L}}
\newcommand{\cP}{{\cal P}}
\newcommand{\cS}{{\cal S}}
\newcommand{\half}{{1\over 2}}
\newcommand{\quart}{{1\over 4}}

\newfont{\Bbb}{msbm10 scaled 1200}     
\newcommand{\mathbb}[1]{\mbox{\Bbb #1}}
\def\IR{{\mathbb R}}

\newcommand{\mysection}[1]{\setcounter{equation}{0}\section{#1}}

\input epsf
\newcounter{fignum}
\newcommand{\figurnum}{\arabic{fignum}}
\newcommand{\figur}[2]{
\addtocounter{fignum}{1}
\addcontentsline{lof}{figure}{\protect
\numberline{\arabic{section}.\arabic{fignum}}{#2}}
\hspace{-3mm}{\it fig.}\ \figurnum.
\begin{figure}[t]\begin{center}
\leavevmode\hbox{\epsffile{#1.eps}}\\[3mm]
\parbox{10cm}{\small \bf Fig.\ \figurnum : \it #2}
\end{center} \end{figure}\hspace{-1.5mm}}
\begin{document}

\begin{titlepage}

\rightline{UCB-PTH-99/11, LBNL-43079, NSF-ITP-99-22}
\rightline{hep-th/9904191}

\begin{center}

\vskip 1.5 cm
{\Large \bf Wilson Loops and Minimal Surfaces}

\vskip 1 cm
{\large Nadav Drukker$^{1,2}$,
David J. Gross$^1$
and Hirosi Ooguri$^{3,4}$}\\

\vskip 1cm
$^1$Institute for Theoretical Physics,
University of California,\\
Santa Barbara, CA 93106

\smallskip
$^2$Department of Physics, Princeton University,\\
Princeton, NJ 08544

\smallskip
$^3$Department of Physics,
University of California,\\
Berkeley, CA 94720-7300

\smallskip
$^4$Lawrence Berkeley National Laboratory, \\
Mail Stop 50A-5101,
Berkeley, CA 94720

\bigskip
{\tt drukker@itp.ucsb.edu,
gross@itp.ucsb.edu,
Hooguri@lbl.gov}

\end{center}

\vskip 0.5 cm
\begin{abstract}
The $AdS$/CFT correspondence suggests that the Wilson loop
of the large $N$ gauge theory with ${\cal N}=4$ supersymmetry
in 4 dimensions is described by a minimal surface in $AdS_5 \times
S^5$. We examine various aspects of this proposal,
comparing gauge theory expectations with computations
of minimal surfaces. There is a distinguished
class of loops, which we call BPS loops, whose expectation
values are free from ultra-violet divergence. We formulate the
loop equation for such loops. To the extent that we have checked,
the minimal surface in $AdS_5 \times S^5$ gives a solution
of the equation. We also discuss the zig-zag symmetry of
the loop operator. In the ${\cal N}=4$ gauge theory, we expect
the zig-zag symmetry to hold when the loop does not couple
the scalar fields in the supermultiplet. We will show how
this is realized for  the minimal surface.

\end{abstract}
\end{titlepage}

\mysection{Introduction}

The remarkable duality between 4-dimensional supersymmetric
gauge
theories and type IIB string theory on $AdS_5\times S^5$
background
\cite{Maldacena:1997re} has been studied extensively over the past year
and a half. This conjecture is difficult to test. As with many
dualities, it relates a weakly coupled string theory to a
strongly coupled gauge theory. Weakly coupled string theory is
well defined, even though there are technical problems in doing
calculations with
Ramond-Ramond backgrounds. But how can one compare the
results to
the gauge theory, which is strongly coupled?
Even if there is no phase transition
in going from weak to strong coupling in the gauge theory, there is
little that can be said about the strongly coupled gauge theory.
By virtue of non-renormalization theorems, it is possible  to calculate
some quantities in perturbation theory and extrapolate to strong
coupling. Such techniques, however, raise the question of whether
these comparisons
can be regarded as  strong evidence for the conjecture or whether
the result
is dictated by symmetry alone.

Gauge theory without fermions has a non-perturbative formulation
on
the lattice.  This allows one to define, if not compute, quantities at
arbitrarily large bare couplings.  The lattice formulation of gauge
theory enables one to derive a rigorous form of the loop equation
\cite{Makeenko:1979pb}, for the large $N$ limit of the theory.  These
equations are
satisfied on the lattice and are solved by the master
field of the theory.  The only case where the loop equation has
been explicitly solved is 2 dimensions, where the theory is soluble
\cite{Kazakov:1980zi}.

The loop equation can also be derived formally
in the continuum field theory. It has been shown that the
perturbative expansion of the theory
yields a solution to the loop equation. This is
also the case for supersymmetric theories. Thus, although there is
no
formulation of supersymmetric theories on the lattice, we
assume that those theories still satisfy a large $N$ loop equation.
Since this equation holds for all couplings we
can use it for strong coupling as well. One of the goals of this paper
is to check if the
$AdS_5$ ansatz for the expectation value of the Wilson loop
operator
satisfies the loop equation.
To the extent that we were able to reliably estimate properties of
string in $AdS_5$,  the loop equation is satisfied.
However we were unable to test them in all interesting cases. In
the course of our investigation we will also learn new facts about
Wilson loops and strings in Anti de-Sitter space.

We discuss the best understood and most studied case of the
$AdS$/CFT
correspondence between type IIB superstring
on $AdS_5\times S^5$ and
$\cN=4$ super Yang-Mills theory
with gauge group $SU(N)$ in 4 dimensions. We
will concentrate on the case with Euclidean signature metric.
Let us review some basic facts about this
duality\footnote{For more complete reviews see
\cite{Schwarz:1998fd,Douglas:1999ww}.}.

The near horizon geometry of $N$ D3-branes is given by the metric
\beq
{ds^2\over \alpha'} ={U^2\over\sqrt{4\pi
g_sN}}\sum_{\mu=0}^3dX^\mu dX^\mu
+\sqrt{4\pi g_sN}{dU^2\over U^2}
+\sqrt{4\pi g_sN}d\Omega_5^2 ,
\eeq
where $g_s$ is the string coupling constant and the
string tension is $(2\pi \alpha')^{-1}$.
The background contains $N$ units of Ramond-Ramond flux.
The $X$ and $U$ are coordinates on $AdS_5$, and
$d\Omega_5^2$ is the metric on $S^5$ with unit radius.
The curvature radii of both $AdS_5$ and $S^5$ are given by
$(4\pi g_sN)^{{1\over 4}} l_s$ where $\alpha' = l_s^2$.
We will find it more convenient to rescale the coordinates $X^\mu$
by
$1/\sqrt{4\pi g_sN}$ and introduce new coordinates
$Y^i=\theta^i/ U$ ($i=1,\cdots,6$),
where $\theta^i$ are the coordinates on $S^5$ and
$\theta^2 = 1$. The metric in this coordinate system is
\beq
{ds^2 \over \alpha'} =\sqrt{4\pi g_s N} Y^{-2}
\left(\sum_{\mu=0}^3dX^\mu dX^\mu+\sum_{i=1}^6dY^i dY^i\right).
\label{metric}
\eeq
It is interesting to note that $AdS_5 \times S^5$ is
conformal to flat $\IR^{10}$ if the radii of $AdS_5$
and $S^5$ are the same.
In this coordinate system, the boundary of $AdS_5$ is mapped to
the
origin $Y^i=0$ of $\IR^6$.

The gauge theory coupling $g_{YM}$ and the string coupling
$g_s$ are related by
$g_{YM}^2=4\pi g_s$.
We are interested in the limit of $N \rightarrow \infty$
while keeping the 't Hooft coupling
$\lambda = g_{YM}^2 N$ finite \cite{'tHooft:1974jz}.
After taking the large $N$ limit, we will consider the
region $\lambda \gg 1$, where the
curvature is small compared to the
string scale and stringy excitations are negligible.
In this case, the supergravity
approximation is reliable. According to the $AdS$/CFT
correspondence, every supergravity field has
a corresponding local operator in
the gauge theory. Correlators of local operators are given by
the supergravity action for fields with point sources
on the boundary of $AdS_5$
\cite{Gubser:1998bc,Witten:1998qj}.
In the classical limit one just solves the equations of motion
with such sources.

An interesting set of non-local operators in a gauge theory are
Wilson loops. It was proposed in
\cite{Maldacena:1998im,Rey:1998ik}
that the Wilson loop is defined by an open string
ending on the loop at the boundary of $AdS_5$. In the classical
limit, the string is described by a minimal surface.
Due to the curvature of $AdS_5$, the minimal surface
does not stay near the
boundary, but goes deep into the interior of space,
where the area element can be made smaller. Because of this the
behavior of the Wilson loop, for large area, is that of a conformal
theory, and the area law does not produce confinement.

The gauge theory under discussion does not contain quarks or
other fields in the fundamental representation of the gauge group.
To construct
the Wilson loop describing the phase associated with moving a
particle in the fundamental representation around a closed curve,
we place one of the D-branes
very far away from the others. The ground states of the
string stretched from the distant D-brane to the others consist of
the $W$-bosons and their superpartners in the fundamental
representation of the gauge group of the
remaining branes. Thus,
for large $\lambda$, the expectation value of the Wilson loop is
related to the classical action of the string, with appropriate
boundary conditions.
To the leading order in $\lambda$, we can ignore the effect
of the Ramond-Ramond flux and  use
the Nambu-Goto action, namely the area of the minimal surface
\beq
A=\int {d\sigma_1d\sigma_2
 \over2\pi\alpha'\sqrt{\lambda}} \sqrt{g}
=\int {d\sigma_1d\sigma_2 \over 2 \pi
Y^2}\sqrt{\det(\partial_\alpha X^\mu\partial_\beta X^\mu+
\partial_\alpha Y^i\partial_\beta Y^i)}.
\label{area}
\eeq
Because of the $Y^{-2}$ factor, this area is infinite. After
regularizing the divergence, the infinite part was identified as  due
to the mass of the $W$-boson and subtracted
\cite{Maldacena:1998im}. Taking 2 parallel
lines (with opposite orientation) as a quark-anti quark pair, the
remaining finite part defines  the quark-anti quark potential.
Such calculations were used to study the phases of the
$\cN=4$ super Yang-Mills theory
and to demonstrate confinement in non-supersymmetric
generalizations \cite{Witten:1998zw,Brandhuber:1998bs}.

We will argue below that the correct action of the Wilson loop is not
the area of the minimal surface, but the Legendre transform of it
with respect to some of the loop variables.
The reason is that some of the string coordinates satisfy Neumann
conditions rather than Dirichlet conditions. For a certain
class of loops, this Legendre transform exactly removes the
divergent piece
from the area. As the result, the expectation values
of such loops are finite.

The appropriate Wilson loop for  $\cN=4$ super Yang-Mills theory
is an operator of the form (suppressing all fermion fields for the
moment)
\beq
W[C] ={1\over N}\Tr\,\cP\,
\exp\left(\oint (iA_\mu \dot x^\mu+\Phi_i\dot y^i)ds\right),
\eeq
where $A_\mu$ are the gauge fields and $\Phi_i$ are the six
scalars in the adjoint representation,
and $C$ represents the loop variables $(x^\mu(s), y^i(s))$.
$(x^\mu(s))$ determines the actual loop in four dimensions, 
$(y^i(s))$ can be thought of as the extra six coordinates of 
the ten-dimensional $\cN=1$ super Yang-Mills theory, of which 
our theory is the dimensionally reduced version.
It turns out that minimal surfaces terminating at the boundary of
$AdS_5$ correspond only to
loops that satisfy the constraint $\dot x^2=\dot y^2$. This
constraint was derived before, and we study in greater depth 
its origin and meaning. In \cite{Maldacena:1998im}, the 
constraint was introduced
as a consequence of the fact that the
mass of the open string and the Higgs VEV are proportional
to each other. We will show that the constraint also has a geometric
interpretation in terms of a minimal surface in $AdS_5 \times S^5$.
Another  interpretation of the constraint has to do with the ${\cal
N}=4$ supersymmetry; the loops obeying
the constraint are BPS-type objects in loop-space. After
discussing various aspects of loops obeying
the constraint,
we present some idea on how to extend the calculation to a more
general class of loops.

The loop equation is a differential equation on the loop space. We
evaluate, using string theory on $AdS_5$, the action of the loop
differential operator $\hat{L}$ on a certain class of Wilson loops.
On a smooth loop $C$, we find the differential operator annihilates
the
vacuum expectation of the loop $\langle W \rangle$, in accord with
the loop equation as derived in the gauge theory. On the other
hand,
for a loop with a self-intersection point, the gauge theory predicts
that $\hat{L} \langle W \rangle$ is non-zero and proportional to
$g_{YM}^2 N$. We point out the gauge theory also predicts
that a cusp (a sharp turning point) in a loop gives
a non-zero contribution to the loop equation, proportional to
$g_{YM}^2 N$. We will show that $\hat{L} \langle W \rangle$
for a loop with a cusp evaluated by the minimal surface
in $AdS_5 \times S^5$ is indeed non-vanishing
and proportional to $g_{YM}^2 N$. We have not been able to
reproduce the precise dependence on the angle at the cusp due
to our lack of detailed understanding of loops not obeying
the constraint $\dot x^2 = \dot y^2$. For the same reason we 
were unable to reproduce the expected result at an intersection.

\medskip
The paper is organized as follows.
\nopagebreak

\nopagebreak
\medskip
\nopagebreak
\noindent
In Section 2, we start with a brief
review of the Wilson loop operator in the pure Yang-Mills theory.
We then point out
an important subtlety in performing the Wick rotation in the
supersymmetric
theory. We will present some results from the perturbation
theory where the subtlety in the Wick rotation plays an interesting
role.

\medskip
\noindent
In Section 3, we turn to string theory in $AdS_5 \times S^5$.
We will give a precise specification of boundary conditions
on the string worldsheet and the geometric
origin of the constraint $\dot x^2 = \dot y^2$.
For some cases, we can compute the area of minimal surfaces
explicitly. These include loops with intersections or
cusps. For such loops, the areas have logarithmic divergences.
After calculating those areas, we explain the need for the
Legendre transform and show that it removes the linear divergence.
The absence of a linear divergence fits well with what we expect
for the supersymmetric gauge theory. We will clarify the issue of
zig-zag symmetry, and end the section with a discussion of
loops that do not satisfy the constraint.

\medskip
\noindent
Section 4, we give a review of the loop equation in the pure 
Yang-Mills theory and derive its generalization to the 
case of ${\cal N} = 4$ super Yang-Mills theory in 4 dimensions.

\medskip
\noindent
In Section 5, we will discuss to what extent the minimal
surface calculation in $AdS_5$ is consistent with the loop equation.

\medskip
\noindent
To make the body of the paper more readable, some details are
presented in appendices. In Appendix \ref{appendix-phase} we
derive the
Wilson loop as the first quantized action of the $W$-boson. In
Appendix \ref{appendix-cusp} we calculate the area of a minimal
surface near a cusp. In Appendix \ref{appendix-N=4}, we present
some
more details on the loop equation of the $\cN=4$ theory.

\mysection{Wilson Loops in N=4 Gauge Theory}

We define the Wilson loop operator in the
supersymmetric gauge theory, and review some of its basic
properties.
We pay particular attention to its coupling to the scalar fields
in the supermultiplet.

\subsection{Definition}
\label{generalities}

One of the most interesting observables in gauge theories is the
Wilson loop, the path-ordered exponential of the   gauge field,
\beq
W={1\over N}\Tr\,\cP\,\exp\left(i\oint A_\mu dx^\mu\right),
\label{wilsonloop}
\eeq
with the trace in the fundamental representation. The Wilson loop
can be defined for any closed path in space, providing
a large class of gauge invariant observables. In fact, these
operators, and their products, form a complete basis of gauge
invariant operators for pure Yang-Mills theory.
An appropriate definition of the loop operator
for the $\cN=4$ super Yang-Mills theory in 4 dimensions will be
given below.

One of physical applications of Wilson loops stems from the fact
that an
infinitely massive quark in the fundamental representation moving
along the loop will be transformed by the phase factor
in (\ref{wilsonloop}).
Thus the dynamical effects of the gauge dynamics on external
quark sources  is measured by the Wilson loop.  In
particular for a parallel quark anti-quark pair, the Wilson loop is
the exponent of the effective potential between the quarks and
serves
as an order parameter for  confinement \cite{Wilson:1974sk}.

The Maldacena conjecture states that
type IIB string theory on
$AdS_5\times S^5$ is  dual to $\cN=4$ super Yang-Mills theory in 4
dimensions. This gauge theory does not contain quarks in the
fundamental representation.
To construct the Wilson loop, we separate a single D-brane from
the $N$ D-branes
and take it very far away. For large $N$, we can ignore the fields
on the distant D-brane, except for open strings stretching between
it and the other $N$. The ground states of the open string
are the $W$-bosons and their superpartners of the broken,
$SU(N)$, gauge group. Their trajectories should
give the same effect as that of an infinitely massive particle in the
fundamental representation.

The correlation functions of the $W$-boson can be written in  the
first quantized formalism as an integral over paths. This  description
is studied in detail
in Appendix \ref{appendix-phase}. When the 4-dimensional space
has the Lorentzian signature metric, the phase factor associated to
the loop is given by the vacuum expectation value of the operator
\beq
W={1\over N}\Tr\,\cP\,
\exp\left(i\oint (A_\mu \dot x^\mu+|\dot x|\Phi_i\theta^i)ds\right).
\eeq
When the metric
is Euclidean,
there is an important modification to this formula as
\beq
W={1\over N}\Tr\,\cP\,
\exp\left(i\oint (A_\mu \dot x^\mu-i|\dot x|\Phi_i\theta^i)ds\right).
\label{euclid-phase}
\eeq
Notice the presence of $i$ in the second term in the exponent. The
``phase-factor'' in the Euclidean theory is not really a phase, but
contains a real part.

In the above, $\theta^i$ are angular coordinates of
magnitude $1$ and can be regarded as coordinates on $S^5$.
In the gauge theory, we may consider a more general class
of Wilson loops of the form
\beq
W={1\over N}\Tr\,\cP\,
\exp\left(\oint (iA_\mu \dot x^\mu+\Phi_i\dot y^i)ds\right).
\label{mostgeneral}
\eeq
with an arbitrary function
$y^i(s)$.  This is the general loop we would get by dimensional
reduction from the 10-dimensional gauge theory, where $\Phi_i$
would be the extra six components  of the gauge field. Equation
(\ref{euclid-phase})
restricts us to the case of $\dot x^2-\dot y^2=0$.
This suggests that the metric on the loop
variables $(x^\mu(s), y^i(s))$ has the signature $(4,6)$.
It is important to stress that this
is not the signature
of $AdS_5\times S^5$ but of the space where the loops
are defined\footnote{One may regard the extra factor
of $i$ in the Euclidean case
(\ref{euclid-phase}) as a Wick-rotation of the 6 $y$-coordinates
so that we can express the
constraint as $\dot x^\mu\dot x_\mu+\dot y^i\dot y_i=0$,
both in the Lorentzian and the Euclidean cases.
To avoid confusions, we will not use this convention and
write the $i$ explicitly in all our expressions in the Euclidean
case.}.
As we will show later, the signature of the loop space metric is
related to the fact that the 6 loop variables $y^i(s)$ correspond
to T-dual coordinates on the string worldsheet.
The constraint $\dot x^2-\dot y^2=0$ is also
related to supersymmetry.

Gauge invariance in 4 dimensions requires that the
Wilson loop close in 4 dimensions, $i.e.$
the loop variables $x^\mu(s)$ are continuous and periodic around
the loop.
This is not the case for the other 6 variables $y^i(s)$,
and the loop may have a jump in these 6 directions.

\subsection{Perturbation Theory}
\label{pertsection}

As a warm-up, we study properties of the Wilson loops in
perturbation theory. To  first order
in $g_{YM}^2 N$,
the expectation value of the loop $\langle W \rangle$ is given by
\ber
\langle W[C] \rangle &=&1-g_{YM}^2N\oint ds\oint ds'
\Big[
\dot x^\mu(s)\dot x^\nu(s') G_{\mu\nu}\left(x(s)-x(s')\right)
\phantom{G_{ij}^2}
\nonumber\\
&&\hskip1.5in
-\dot y^i(s)\dot y^j(s') G_{ij}\left(x(s)-x(s')\right)
\Big],
\eer
where $G_{\mu\nu}$ and $G_{ij}$ are the gauge field and scalar
propagators. The relative minus sign comes from the extra $i$ in
front of the scalar piece in the exponent in (\ref{euclid-phase}).
This integral is linearly divergent.  With a regularization of
the propagator with cutoff $\epsilon$ ({\it i.e.} replacing $1/x^2$ with
$1/(x^2+\epsilon^2)$), the divergent piece
coming from the exchange of the gauge field $A_\mu$ is evaluated
as
\beq
-{\lambda\over8\pi^2}
\oint ds\int_{-{\epsilon\over|\dot x|}}^{\epsilon\over|\dot x|}ds'
\dot x^\mu(s)\dot x^\nu(s') {\delta_{\mu\nu}\over\epsilon^2}
=-{\lambda\over(2\pi)^2\epsilon}\oint ds|\dot x|
=-\lambda{L\over(2\pi)^2\epsilon},
\label{pert-A}
\eeq
where $L$ is the circumference of the loop. The divergent
contribution
from the exchange of the scalars $\Phi_i$ is
\beq
{\lambda\over8\pi^2}
\oint ds\int_{-{\epsilon\over|\dot x|}}^{\epsilon\over|\dot x|}ds'
\dot y^i(s)\dot y^j(s') {\delta_{ij}\over\epsilon^2}
={\lambda\over(2\pi)^2\epsilon}\oint ds |\dot x|
{\dot y^2\over \dot x^2}.
\label{pert-Phi}
\eeq
Combining these terms together, we find
\beq
  W = 1 + {\lambda \over (2\pi)^2 \epsilon} \oint ds |\dot x|
\left(1 - {\dot y^2 \over \dot x^2} \right) + {\rm finite}.
\label{pert-div}
\eeq
We note that  the linear divergence cancels when
the constraint $\dot x^2=\dot y^2$ is satisfied.

At $n$-th order in the $\lambda = g_{YM}^2 N$ expansion,
one finds a linear divergence of the form
\beq
{\lambda^n\over\epsilon}
\oint ds|\dot x|G_n \left({\dot y^2 \over \dot x^2}\right),
\eeq
for some polynomial $G_n(z)$. We now argue that $G_n(1)=0$,
namely the linear divergence cancels when $\dot x^2 = \dot y^2$,
to all order in the perturbative expansion.
The $n$-th order term is calculated by connected Feynman
diagrams
with external legs attached to the loop.
The linear divergence appears when all the external legs come
together
in 4 dimensions. Since the Feynman rule
of the ${\cal N}=4$ gauge theory is obtained by the dimensional
reduction of the 10-dimensional theory,
the 10-dimensional rotational invariance of the Feynman rule
is recovered in the coincidence limit.
Therefore the contractions of the external indices by
the Feynman rule produce only rotational
invariant combinations of $(\dot x^\mu, i \dot y^i)$, namely a
polynomial of $(\dot x^2 - \dot y^2)$. The polynomial does not
have a constant term since a connected Feynman diagram for
$\langle W \rangle$ needs to have at least 2 external lines
attached to  the loop. Therefore the polynomial vanishes when
$\dot x^2 - \dot y^2=0$.

When the loop has a cusp, there is an extra logarithmic divergence
from graphs as shown in \figur{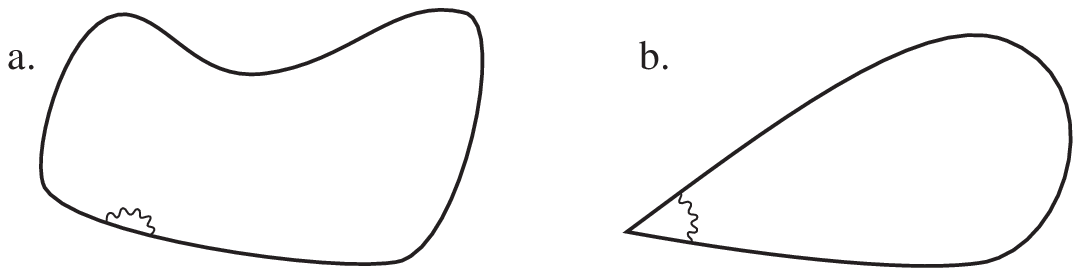}
{(a) At one loop, there is a linear divergence from the propagator
connecting coincident points. The divergence is proportional to the
circumference of the loop. (b) At cusps and intersections, an
additional logarithmic divergence appears when the 2 external
legs approach the singular point.}
Let us denote the angle at the cusp by $\Omega$.
We choose the angle so that $\Omega = \pi$ at a regular point
of the loop. A one-loop computation
with the gauge field gives
\beq
{\lambda\over(2\pi)^2}((\pi-
\Omega)\cot\Omega+1)\log{L\over\epsilon}.
\label{oneloop1}
\eeq
A cusp is a discontinuity of $\dot x^\mu$. There may also be a
discontinuity in $\dot y^i$, which we measure by an angle $\Theta$.
We choose $\Theta$ so that
$\Theta=0$ when $\dot y^i$ is continuous.
A one-loop computation with the scalar fields gives
\beq
-{\lambda\over (2\pi)^2}\left(-{\pi-
\Omega\over\sin\Omega}\cos\Theta+1\right)
\log{L\over\epsilon}.
\label{oneloop2}
\eeq
Combining (\ref{oneloop1}) and (\ref{oneloop2}) together, we
obtain
\beq
{\lambda\over (2\pi)^2}{\pi-
\Omega\over\sin\Omega}(\cos\Omega+\cos\Theta)
\log{L\over\epsilon}.
\label{pert-log}
\eeq
A similar computation at an intersection gives
\beq
{\lambda\over2\pi}{1\over\sin\Omega}(\cos\Omega+\cos\Theta)
\log{L\over\epsilon}.
\label{pert-int}
\eeq

\mysection{Minimal surfaces in Anti de-Sitter Space}
\label{section-minimal}

According to the Maldacena conjecture, the expectation value of the
Wilson loop is given by the action of a string bounded by the curve
at the boundary of space:
\beq
\langle W[C]\rangle
=\int_{\partial X=C}\cD X\, \exp( -\sqrt{\lambda} S[X]),
\label{pathintegral}
\eeq
for some string action $S[X]$. Here $X$ represents both the
bosonic
and the fermionic coordinates of the string. For large $\lambda$,
we can estimate the path integral by the steepest descent method.
Consequently the expectation value of the Wilson loop
is related to the area $A$ of the minimal surface
bounded by $C$ as
\beq
\langle W \rangle \simeq \exp ( - \sqrt{\lambda} A ) .
\label{firstansatz}
\eeq
The motivation for this ansatz is that the $W$-boson considered in
section \ref{generalities} is described in the D-brane language by an
open string going between the single separated D-brane and the
other
$N$ D-branes. In the near horizon limit, the $N$ D-branes are
replaced by the $AdS_5$ geometry and the open string is stretched
from the boundary to the interior of $AdS_5$.

To be precise, this argument only tells us that the Wilson loop and
the string in $AdS_5$ are related to each other. The expression
(\ref{pathintegral}) is schematic at best, and there may be
an additional loop-dependent factor in (\ref{firstansatz}).
A similar problem exists in computation of correlation functions of
local operators; there is no known way to fix
the relative normalization of local operators in the gauge theory
and supergravity fields in $AdS_5$. To determine the normalization
factor, one has to compute the 2 point functions
\cite{Freedman:1998tz,Lee:1998bx}.
In our case, the normalization factor in (\ref{firstansatz})
may depend on the loop variables $C=(x^\mu(s), y^i(s))$.
In fact, we will argue below that the correct
action to be used in (\ref{firstansatz}) is not the area $A$
of the surface, but the  Legendre transform of it. This modification
does not change the equations of motion, and the solutions are still
minimal surfaces. However the values of the classical action for
these  surfaces are different than their areas.

We will assume that, to the leading order in $\lambda$,
there is no further $C$ dependent factor.
Otherwise the conjecture would be meaningless as it would
produce no falsifiable predictions.
On the other hand, one expects a $C$ dependent factor
in the subleading order, such as the fluctuation determinant
of the surface in $AdS_5$. There can also be a factor in
the relation between the $W$-boson propagation amplitude and the
Wilson
loop computed in Appendix A. Such a factor would be kinematic in
nature
and independent of $\lambda$, and therefore negligible in
our analysis.

\subsection{Boundary Conditions and BPS Loop}

The Wilson loop discussed in \cite{Maldacena:1998im}
obeys the constraint
\beq
\dot x^2=\dot y^2.
\label{constraint}
\eeq
This constraint was originally derived by using the coupling of the
fundamental string to the gauge fields and to the scalars.
In our derivation of the loop operator from the phase factor
for the $W$-boson amplitude in Appendix \ref{appendix-phase},
the constraint arises from the saddle point
in integrating over different reparametrizations of the same loop;
essentially for the same reason as in \cite{Maldacena:1998im}.

In this section, we will give another interpretation of the constraint
(\ref{constraint}), in terms of the string theory in $AdS_5 \times
S^5$.
For this interpretation we need to give a precise specification of the
boundary
condition on the string in $AdS_5 \times S^5$.

We begin with   super Yang-Mills theory in 10 dimensions,
which is realized on space-filling D$9$-branes. We ignore
the fact that this theory is anomalous
since we will
reduce it to the anomaly free theory in 4 dimensions.
Moreover, we are only
interested in the boundary conditions on bosonic
variables\footnote{Boundary conditions for fermionic variables are
not relevant in our analysis of the loop for  large $\lambda$.}.
The Wilson loop in 10 dimensions corresponds to an open string
worldsheet bounded by the loop, $i.e.$ we should
impose full Dirichlet
boundary conditions on the string worldsheet.
This is natural since, without the Wilson loop operator,
the string end-point obeys fully Neumann boundary conditions
along the D$9$-brane. The conditions imposed by the
Wilson loop are complementary to the boundary conditions
on the D$9$-brane.

To reduce the theory to 4 dimensions, we perform T-duality
along 6 directions. An open string ending on the D$3$-brane
obeys 4 Neumann and 6 Dirichlet boundary conditions.
Consequently, the Wilson loop operator in the 4-dimensional gauge
theory
imposes complementary boundary conditions; namely 4 Dirichlet
and 6 Neumann boundary conditions. If the Wilson loop is
parametrized
by the loop variables $(x^\mu(s), y^i(s))$, where   $\dot y^i(s)$
couples
to the 6 scalar fields, then the 6 loop variables $\dot y^i(s)$
are to be identified with
the 6 Neumann boundary conditions on the string worldsheet.

We are ready to specify the boundary condition on the string
worldsheet living in
$AdS_5 \times S^5$, with line element,
\beq
{ds^2 \over \alpha'} = \sqrt{\lambda} Y^{-2}
\left(\sum_{\mu=0}^3dX^\mu dX^\mu+\sum_{i=1}^6 dY^i dY^i\right).
\eeq
Choose the string world-sheet coordinates to be $(\sigma^1,
\sigma^2)$ such that the boundary is located at $\sigma^2=0$.
Since $X^\mu$ is identified with the   4 dimensional coordinates
where the gauge theory lives, it is natural to impose
Dirichlet conditions on $X^\mu$, so that
\beq
X^\mu(\sigma_1,0)=x^\mu(\sigma_1).
\label{dirichlet}
\eeq
The remaining 6 string coordinates $Y^i(\sigma^1, \sigma^2)$
obey Neumann boundary conditions. We propose
that these boundary  conditions are
\beq
 J_1^{~\alpha} \partial_\alpha Y^i(\sigma^1,0)
 = \dot y^i(\sigma^1),
\label{neumann}
\eeq
where $J_\alpha^{~\beta}$ ($\alpha,\beta = 1,2$) is the complex
structure on the string worldsheet given in terms of the induced
metric
$g_{\alpha\beta}$,
\beq
   J_\alpha^{~\beta} = {1 \over \sqrt{g}} g_{\alpha\gamma}
  \epsilon^{\gamma\beta}.
\eeq

Although we do not have a derivation of the boundary condition
(\ref{neumann}) from first principles, it can be motivated
as follows. Because of the identification of the $SO(6)$ symmetries
in the $AdS$/CFT correspondence, it is clear that Neumann
boundary
conditions must set $\dot y^i$ equal to $J_1^{~\alpha}
\partial_\alpha
Y^i$ up to a relative normalization of the two.
The use of the induced complex structure $J_\alpha^{~\beta}$
in the Neumann boundary condition
is required by the reparametrization invariance on the worldsheet.
The fact that the condition
$\dot x^2 = \dot y^2$ has a natural interpretation in
terms of the minimal surface, as we will explain below,
suggests that the normalization factor is $1$, as in
(\ref{neumann}).

For a generic choice of the loop variables $(x^\mu(s), y^i(s))$,
there is a unique minimal surface in Euclidean space obeying
the 10 boundary conditions,
(\ref{dirichlet}) and (\ref{neumann}).
However the resulting minimal surface does not necessarily
terminate
at  the boundary $Y^i=0$ of $AdS_5$.  The condition
$Y^i=0$ would be additional Dirichlet conditions, which
may or may not be compatible with (\ref{neumann}). In fact,
one can show that, for a smooth loop, the additional condition
$Y^i(\sigma^1,0)=0$ is satisfied by the minimal
surface if and only if the loop variables obey
the constraint $\dot x^2 = \dot y^2$.
To see this consider the Hamilton-Jacobi
equation\footnote{
In general, the Hamilton-Jacobi equation for the area of a minimal
surface
on a Riemannian manifold with a metric $G_{IJ}$ takes the form,
$$G^{IJ} (\delta A/\delta X^I) (\delta A/\delta X^J)
 = G_{IJ} \partial_1 X^I \partial_1 X^J.$$}
for the area $A$ of a minimal surface
bounded by a loop $(X^\mu(s), Y^i(s))$ in $AdS_5 \times S^5$:
\beq
   \left( {\delta A \over \delta X^\mu} \right)^2
  + \left( {\delta A \over \delta Y^i} \right)^2
 =  {1 \over (2 \pi )^2Y^4} \left( (\partial_1 X^\mu)^2
  + (\partial_1 Y^i)^2 \right).
\eeq
Since the momenta conjugate to the $X^\mu$'s and the $Y^i$'s are
given by
\beq
 {\delta A \over \delta X^\mu} =
{1 \over 2\pi Y^2} J_1^{~\alpha} \partial_\alpha
 X^\mu, \qquad {\delta A \over \delta Y^i}
      = {1 \over 2 \pi Y^2} J_1^{~\alpha} \partial_\alpha
 Y^i,
\eeq
we obtain
\beq
  (J_1^{~\alpha} \partial_\alpha X^\mu)^2
  + (J_1^{~\alpha} \partial_\alpha Y^i)^2
  = (\partial_1 X^\mu)^2 + (\partial_1 Y^i)^2.
\eeq
If the minimal surface obeys the boundary conditions
(\ref{dirichlet}) and (\ref{neumann}), this becomes
\beq
  \dot x^2 - \dot y^2
= (J_1^{~\alpha} \partial_\alpha X^\mu)^2
  - (\partial_1 Y^i)^2.
\label{hamjac}
\eeq
Now impose the additional condition that the string worldsheet 
terminates at the boundary of $AdS_5$, $i.e.$ $Y^i(\sigma^1,0)=0$.  
Obviously $\partial_1 Y^i(\sigma^1,0)=0$.  This alone tells us that 
$\dot x^2 - \dot y^2 \geq 0$.  Moreover, if the boundary is smooth, 
it costs a large area to keep 
$J_1^{~\alpha} \partial_\alpha X^\mu$ non-zero near the 
boundary of $AdS_5$, so it has to vanish at the boundary $Y=0$
\cite{Maldacena:1998im}.  Therefore, the condition 
that the minimal surface terminates at the boundary of $AdS_5$ 
requires $\dot x^2 = \dot y^2$.

When the constraint $\dot x^2 = \dot y^2$ is satisfied,
one can reinterpret the 6 Neumann condition (\ref{neumann})
as Dirichlet conditions on $S^5$. To see this, it is useful
to decompose the 6 coordinates $Y^i$
as
\beq
     Y^i = Y \theta^i
\eeq
where $\theta^i$ are coordinates on $S^5$ and
$Y = U^{-1}$ is one of the coordinates on $AdS_5$.
Since for a smooth loop the classical solution has
$\partial_\alpha Y^i = (\partial_\alpha Y) \theta^i$
at the boundary $Y=0$ of $AdS_5$, the Neumann conditions
(\ref{neumann}) turn into the Dirichlet conditions on
$S^5$ as
\beq
   \theta^i(\sigma^1,0) = {\dot y^i \over |\dot y|}.
\label{fixedpt}
\eeq
This justifies
 the boundary conditions used in \cite{Maldacena:1998im}.

There is yet another interpretation of
the constraint $\dot x^2 = \dot y^2$, and it has
to do with supersymmetry.
The loops we have considered so far couple only to bosonic fields:
the gauge field $A_\mu$ and scalars $\Phi^i$.
We also need to allow coupling to
the fermionic fields in the exponent. Fermionic
variables $\zeta(s)$ along the loop couple to the
gauginos $\Psi$ as
\beq
\bar\zeta(\dot x^\mu\Gamma_\mu-i\dot y^i\Gamma_i)\Psi,
\eeq
where we are
using 10-dimensional gamma matrices $\Gamma_\mu$ and
$\Gamma_i$
with signature (10,0). This
is derived in Appendix \ref{appendix-N=4}.
Exactly when the constraint is satisfied this combination of
gamma matrices becomes nilpotent. Consequently only half the
components of $\zeta$ couple to $\Psi$, putting the loop in
a short representation of local supersymmetry in 
super loop-space.
The simplest example is when the Wilson loop is a straight line,
when $\dot x$ and $\dot y$ are independent of $s$. If
$\zeta$ is also constant, this loop is the phase factor associated
with the  a trajectory of a free
BPS particle.

\subsection{Calculating the Area}
\label{section-area}

The computation of the Wilson loop in $AdS_5$ requires an infrared
regularization, since
the area of the minimal surface terminating at the boundary
of $AdS_5$ is infinite due to the factor $Y^{-2}$ in the metric.
In order to make sense of the ansatz (\ref{firstansatz}),
we need to regularize the area.  One natural way to do
so is to impose the boundary conditions (\ref{dirichlet})
and (\ref{neumann}) at $Y=0$, but
integrate the area element only over the part of the surface with
$Y \geq \epsilon$.
On the gauge theory side, the Wilson loop requires
regularization in the ultraviolet.
According to the UV/IR relation in the $AdS$/CFT correspondence
\cite{Susskind:1998dq},
the IR cutoff $\epsilon$ in $AdS_5$ should be identified with the UV
cutoff in the gauge theory.

\medskip

There are a few cases when minimal surfaces can be studied
analytically.

\smallskip
\noindent
(1) Parallel Lines:

\noindent
The minimal surface for parallel lines, each of length $L$ and
separated by a distance $R$, was obtained
in \cite{Maldacena:1998im,Rey:1998ik}. The area of the loop is
\beq
A={2L\over2 \pi \epsilon}
-{4\pi\sqrt{2}\over\Gamma(1/4)^4}{L\over R}.
\label{parallel}
\eeq

\smallskip
\noindent
(2) Circular Loop:

\noindent
The minimal surface in $AdS_5$ bounded by a circle of radius $R$
is found in \cite{dgo,Berenstein:1998ij} as
\beq
Y(r,\varphi)=\sqrt{R^2-r^2},
\label{circlesol}
\eeq
where $r$ and $\varphi$ are radial coordinates on a plane
in the 4 dimensions,
and we use them as coordinates on the string worldsheet also.
The area of the surface with the cutoff $\epsilon$ is
\beq
A={1\over2\pi}\int dr \,r d\varphi\,Y^{-2}\sqrt{1+Y^{\prime2}}
=R\int_0^{\sqrt{R^2-\epsilon^2}}{r\,dr\over(R^2-r^2)^{3\over2}}
={2\pi R\over 2 \pi \epsilon}-1.
\label{circle}
\eeq

\smallskip
\noindent
\nopagebreak{
(3) Cusp:

\noindent
}
Another family of minimal surfaces we can solve analytically
is a surface near a cusp on $\IR^4$ and its
generalization including a jump on $S^5$. We can
find analytical solutions in this case since
the boundary conditions are scale invariant.
Using radial coordinates in the vicinity of  the cusp, $r$ and
$\varphi$, as world sheet coordinates, the scale invariant ansatz,
\beq
Y(r,\varphi)={r\over f(\varphi)},
\eeq
reduces the determination of the minimal surface to a 
one-dimensional problem.
The resulting surface is depicted in \figur{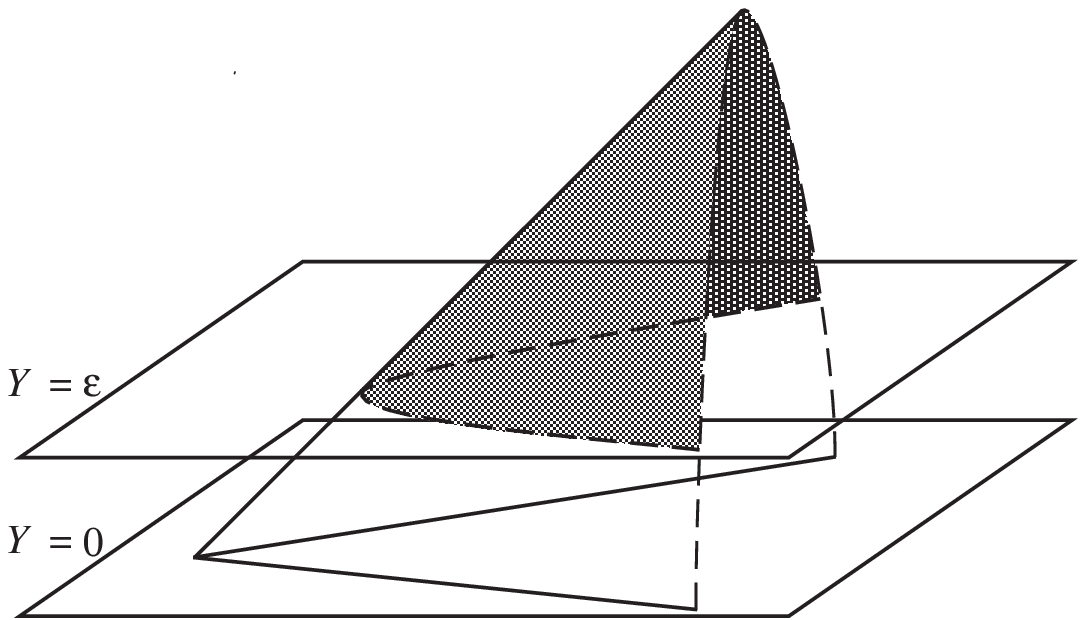}{A minimal
surface for a Wilson loop with a cusp. The regularized area
is evaluated over the shaded region.}
When there is also a jump on $S^5$, one needs to introduce
another variable.
An analytical solution in this case is found in a similar way.
These solutions are presented in Appendix \ref{appendix-cusp}.
The
result is that the area of the surface has a logarithmic
divergence as well as a linear divergence. It behaves as
\beq
A={L\over2 \pi \epsilon}
-{1\over2\pi}F(\Omega,\Theta)\log{L\over\epsilon} + \cdots,
\label{cusp}
\eeq
where $\Omega$ and $\Theta$ are the cusp angles in $R^4$ and
$S^5$
respectively.

When either $\Theta$ or $\Omega$ vanishes,
we can express $F(\Omega,\Theta)/2\pi$
in terms of elliptic integrals.
In \figur{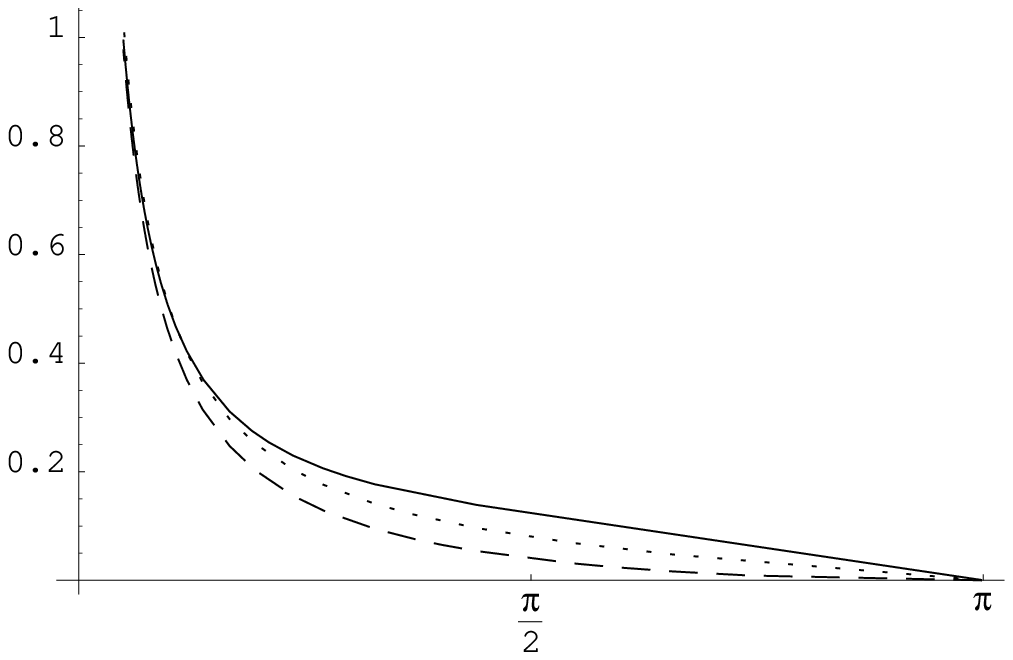}{The solid curve shows the
function $F(\Omega,0)/2\pi$, which appears
in the logarithmic divergence of the minimal
surface with the cusp of angle $\Omega$. This
is compared with the perturbative
result (\ref{pert-log}) at the cusp shown in the dashed curve.
The dotted curve is half of the perturbative result (\ref{pert-int})
at an intersection.}
we show the numerical evaluation of
the function $F(\Omega, 0)$ in the solid curve. This is
to be compared with the perturbative expression
(\ref{pert-log}) shown in the dashed curve.
The function $F(\Omega, 0)$ is zero at $\Omega=\pi$ and has a
pole at $\Omega=0$. As the angle $\Omega \rightarrow 0$ at the
cusp,
the loop goes back along
it's original path, or backtracks. Regularizing the extra divergence
from the pole turns it into a linear divergence which cancels part of
the linear divergence from the length of the loop. This is related to
issues discussed in the section on the zig-zag symmetry.

Away from the cusp, the surface approaches the boundary
along the $Y$-direction without a momentum in the $X$-direction.
Right at the cusp, however, the surface
has momentum in both the $Y$ and
$r$ direction. This means that, although the constraint
 $\dot x^2 = \dot y^2$ is obeyed almost everywhere,
it is modified at the cusp as
\beq
\dot x^2= \big( 1 +f_0^2 \big) \dot y^2,
\label{bad-cusp}
\eeq
where $f_0=f(\varphi=\Omega/2)$ is the minimal value of
$f(\varphi)$.

\medskip
\noindent
(4) Intersection:

\noindent
The minimal surface for a self-intersecting loop is just the sum
of 2 cusps. The only difference is that, by the exchange symmetry
of
the 2 components of the loop, the intersection
forces
\beq
{\dot y\over|\dot x|} = 0
\label{bad-intersection}
\eeq
 instead of (\ref{bad-cusp}).

\medskip

In all the examples above,  there
is a linear divergence $(2\pi \epsilon)^{-1}$ in the regularized
area. This is true for any loop. As explained in
\cite{Maldacena:1998im}, this
leading divergence in the area of the minimal surface in $AdS_5$ is
proportional to the circumference of the loop\footnote{
We are using the coordinates $X^\mu$ in (\ref{metric}) 
to describe the configurations of the Wilson loops. 
With these coordinates, there is no factor of $\lambda$
in the relation between the IR cutoff $\epsilon$ in $AdS_5$ 
and the UV cutoff of the gauge theory \cite{Susskind:1998dq}. 
These coordinates are different from the coordinates on the 
D$3$-brane probe, by a factor of $\sqrt{\lambda}$ \cite{Peet:1999wn}.}.  
The linear divergence
arises from the leading behavior of the surface at small $Y$,
$i.e.$ near the boundary of $AdS_5$.

In this section, we have computed the regularized area by imposing
the boundary condition at the boundary $Y=0$ of $AdS_5$
and integrating the area element
over the part of the surface $Y \geq \epsilon$.
This is not the  unique way to regularize the area.
Another reasonable way to compute the minimal
surface is
to impose the boundary conditions, not at $Y=0$, but
at $Y = \epsilon$. The area bounded by the loop on
$Y=\epsilon$ is then by itself finite.
A comparison of the two regularization prescriptions are
illustrated in \figur{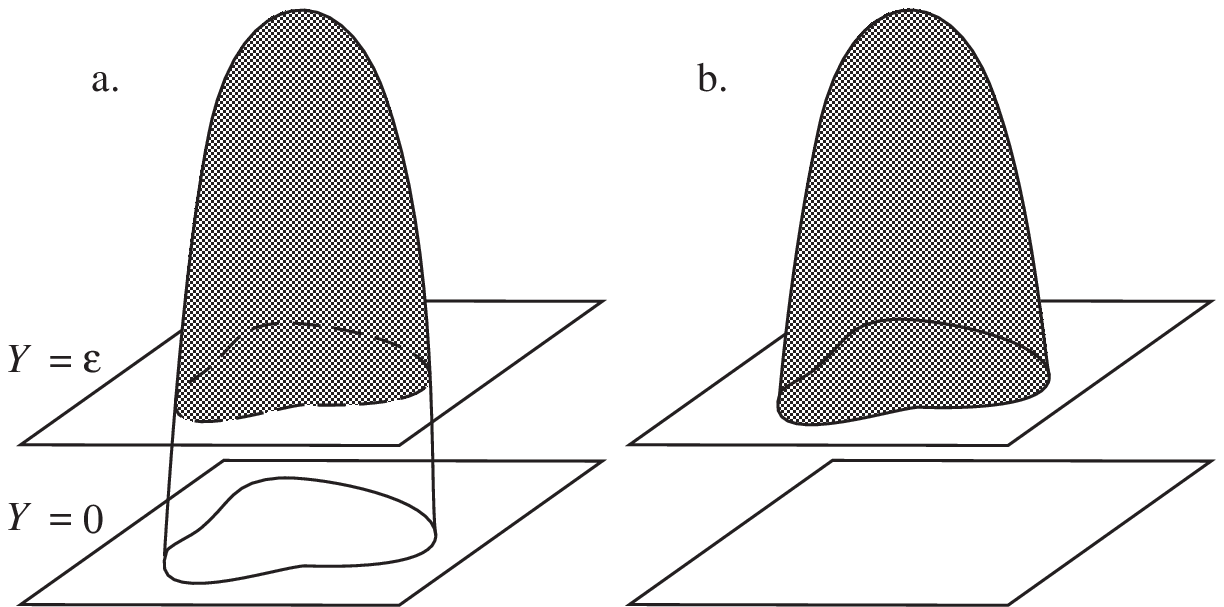}{The comparison of the two regularization
prescriptions. The boundary conditions are imposed
at $Y=0$ in (a) and at $Y=\epsilon$ in (b). The shaded
regions represent the regularized areas.}
These two regularizations give
the same values for the area, up to terms which vanish
as $\epsilon \rightarrow 0$. For example, consider the
circular loop. The solution (\ref{circlesol}) can also be
regarded as a minimal surface with the boundary condition
on $Y=\epsilon$, except that the radius of the circle
on $Y=\epsilon$ is now $R_0=\sqrt{R^2-\epsilon^2}$.
The area computed in this new regularization is then
\beq
A
={1\over\epsilon}\sqrt{R_0^2+\epsilon^2}-1
={2\pi R_0\over2 \pi\epsilon}-1+{\epsilon\over 2R_0}+\ldots.
\eeq
Thus the results of the two regularizations are the
same up to terms which vanish as $\epsilon \rightarrow 0$.
It is straightforward to show that this is also
the case for the parallel lines.
We have also verified that when the loop has a cusp
or an intersection,
the two regularizations give the same area modulo terms
which are finite as $\epsilon \rightarrow 0$, which are
subleading compared to the logarithmic divergence.

When we impose the boundary condition at $Y=\epsilon$,
the constraint on the loop variables is not exactly
$\dot x^2 = \dot y^2$, but it is modified. If the loop
is smooth, the modification is only by $O(\epsilon)$
terms\footnote{
If the loop has a cusp or an intersection, as we saw
earlier, the boundary conditions imposed at $Y=0$
imply the constraint
$\dot x^2=\dot y^2$ holds almost everywhere along the loop,
except at a cusp or an intersection point.
When we impose the boundary
conditions at $Y=\epsilon$, the constraint is modified
in regions of size $\epsilon$ near the cusp
and the intersection point.}.
Therefore most of the results in this paper are independent of
the choice between the two ways of imposing the boundary
conditions.  The only exception
to this rule is the discussion of the zig-zag symmetry.
The zig-zag symmetry of the string worldsheet on
$AdS_5$ seems to fit well with our expectations about
the gauge theory when we use the boundary conditions
at $Y=\epsilon$ rather than at $Y=0$.

\subsection{Legendre Transformation}
\label{corrected-section}

The Maldacena conjecture implies that the Wilson loop is
related to a string ending along the loop on the boundary of space.
In the classical limit, we expect that the string worldsheet
is described by a minimal surface.
This argument, however, does not completely  determine the value
of $\langle W \rangle$ for  large $\lambda$ since there are
many actions whose equations of motion are solved by minimal
surfaces. They differ by total derivatives, or boundary terms. Since 
the surface has boundaries, such terms can be important. 
In \cite{Maldacena:1998im,Rey:1998ik}
it was assumed that one should use the Nambu-Goto action, so the
Wilson loop was given in terms of
the area $A$ of the minimal surface. This is
what we have studied so far. In this section, we argue that
$\langle W \rangle$ is in fact given not by $A$ but by an appropriate
Legendre transform.

We have shown that the loop variables $\dot y^i$
impose Neumann boundary conditions (\ref{neumann})
on the coordinates $Y^i$.
Therefore $\langle W \rangle$ should be regarded as a functional of
the
coordinates $X^\mu$ and the momenta $P_i$ conjugate
to $Y^i$, defined by,
\beq
P_i
={\delta A\over\delta\partial_2Y^i}
={1\over2\pi\sqrt{\lambda} \alpha'}
\sqrt{g}g^{2\alpha}\partial_\alpha Y^j G_{ij}.
\eeq
The Nambu-Goto action is a natural functional of $X^\mu(s)$
and $Y^i(s)$ and is more appropriate for the full
Dirichlet boundary conditions. To replace it
with a functional of $X^\mu(s)$ and $P^i(s)$, we need to perform the
Legendre
transform
\beq
\tilde\cL
=\cL- \partial_2\left(P_i Y^i\right),
\eeq
or
\beq
\tilde A
=A-\oint d\sigma_1 P_i Y^i.
\label{new-action}
\eeq
To show that $\tilde A$ is a natural functional of $(X^\mu,P^i)$,
we use Hamilton-Jacobi theory.
Under a general variation of the $Y$ coordinates,
 the variation of the area $A$ of the minimal surface
is given by
\ber
\delta A
&=&
\int d\sigma_1d\sigma_2
\left({\delta A\over\delta Y^i}
-\partial_\alpha{\delta A\over\delta\partial_\alpha Y^i}
\right)
\delta Y^i(\sigma_1,\sigma_2)
+\oint d\sigma_1{\delta A\over\delta\partial_2 Y^i}\delta
Y^i(\sigma_1,0)
\nonumber\\
&=&
\oint d\sigma_1P_i(\sigma_1,0)\delta Y^i(\sigma_1,0).
\eer
Here we used the equations of motion. Therefore,
after performing the Legendre transformation, we obtain
\beq
\delta\tilde A
=-\oint d\sigma_1Y^i(\sigma_1,0)\delta P_i(\sigma_1,0).
\eeq
Thus $\tilde A$ is a functional of the momenta $P^i$ at the
boundary, not the coordinates $Y^i$.

The Neumann boundary conditions (\ref{neumann})
are conditions on the momenta $P^i$,
\beq
{\dot y^i\over2\pi}=P^i= Y^2 P_i.
\eeq
In fact, if the loop variables $\dot y^i(s)$ are continuous,
the coordinates $Y^i$ are parallel to the momenta $P_i$,
as we saw in (\ref{fixedpt}). In this case, the Legendre
transform gives
\beq
\tilde A
=A-{1\over2\pi}\oint d\sigma_1 {\dot y^i \over Y^2} Y^i
=A-{1\over2\pi}\oint d\sigma_1 {|\dot y|\over Y}
=A-{1\over2\pi\epsilon}\oint ds|\dot y|,
\label{legendre2}
\eeq
where $\epsilon$ is the regulator. In the last step,
we have set $Y=\epsilon$ since the regularized
action is evaluated for $Y \geq \epsilon$.

In the previous section, we saw that the area $A$ of minimal
surface
has a linear divergence proportional to the circumference of the
boundary. By combining it with (\ref{legendre2}),
we find
\beq
\tilde A
={1\over2\pi\epsilon}\oint ds\left(|\dot x|-|\dot y|\right)
+\hbox{finite}
\eeq
for a smooth loop. Therefore the linear divergence cancels
when the constraint
$\dot x^2 = \dot y^2$ is satisfied. The minimal surface
in $AdS_5$ is supposed to describe the Wilson loop for
large coupling $\lambda$. We saw in section
(\ref{pertsection}) that the cancellation of the
divergence also takes place to all order in the perturbative
expansion $\lambda$. This suggests that the cancellation of
the linear divergence is exact, and
a smooth loop obeying $\dot x^2 = \dot y^2$ does not
require regularization. We suspect that this is a consequence
of the BPS property of the loop.
When the loop is a straight line, it preserves a global 
supersymmetry, not only the local one. In that case the lowest order 
perturbation calculation is exact. The modified action is zero, the 
expectation value of the Wilson loop is 1.

We were not able to find an explicit expression for $\tilde\cL$ as a
function of
$X^\mu$, $P^i$ and their derivatives. We only know how to
evaluate it for classical solutions in terms of
the old variables.

By definition, the area $A$ of the minimal surface is positive.
On the other hand, its Legendre transform $\tilde A$
may be negative and the expectation value of the
loop $\langle W \rangle = \exp(-\sqrt{\lambda} \tilde{A})$
may be larger than $1$.
In the pure Yang-Mills theory, the Wilson loop
is a trace of a unitary operator (divided by the rank $N$ of the
gauge group),  and its expectation value has to obey
the inequality $\langle W \rangle \leq 1$.
This is not the case in the supersymmetric theory in the Euclidean
signature space since $W$ in (\ref{euclid-phase}) is not a pure
phase, and there is no unitarity bound on its expectation
value.

We have shown that the expectation value of
a smooth Wilson loop obeying $\dot x^2 = \dot y^2$
is finite.
If the loop has a cusp or an intersection,
the cancellation is not exact and we are left with
the logarithmic divergence.\footnote{
If $\Theta\neq0$, the function $F(\Omega,\Theta)$ gets a
contribution
from the Legendre transformation.}
\beq
\tilde A = - {1\over2\pi} F(\Omega,\Theta)
\log{L\over\epsilon}
+\hbox{finite}.
\eeq
It is interesting to note that the constraint
$\dot x^2 = \dot y^2$ is not satisfied either
at a cusp
\beq
      \dot x^2 = \big(1 + f_0^2  \big) \dot y^2,
\eeq
or at an intersection point
\beq
  {\dot y^i\over|\dot x|} = 0
\eeq
We suspect that the logarithmic divergences at
the cusp and the intersection are
caused by the failure of the loop to satisfy
the BPS condition at these points.

\subsection{Zig-Zag Symmetry}

A Wilson loop of the form
\beq
W={1\over N}\Tr\,\cP\,\exp( i\oint ds A_\mu \dot x^\mu)
\eeq
is reparametrization invariant, in $s$, namely unchanged by 
$s\to f(s)$. Formally it is even invariant under
reparametrizations which backtrack (namely when $\dot f(s)$ is not
always positive) since the phase factor  going forward
and then backwards will cancel. Polyakov has argued in
\cite{Polyakov:1997tj} that this \lq\lq zig-zag symmetry"
is one of the basic properties of the  QCD string.
One must however be careful, even in pure Yang-Mills theory, since
the loop requires regularization. Zig-zag symmetry, in fact, is only
true perturbatively for regularized loops, where the backtracking
paths are closer than the ultraviolet cutoff.
It was pointed out in \cite{Maldacena:1998im} that the Wilson loop
in the supersymmetric theory (\ref{euclid-phase}), with
the constraint $\dot x^2 = \dot y^2$, does not have this symmetry.
This is because the couplings of the Wilson loop to the scalar fields
$\Phi^i$ is proportional to $|\dot x|$, which does not change the
sign when the loop backtracks. Thus if the loop stays at
the same point $\theta^i$ on $S^5$, there is no cancellation of the
coupling to the scalar fields.

In perturbation theory, one can easily prove that
the zig-zag symmetry holds for the Wilson loop (\ref{mostgeneral})
when $\dot y^i = 0$. Suppose we have a segment $C_1$ of a loop
which goes
in one direction and another segment $C_2$ which comes back
parallel
to $C_1$ but in the opposite direction, as shown in
\figur{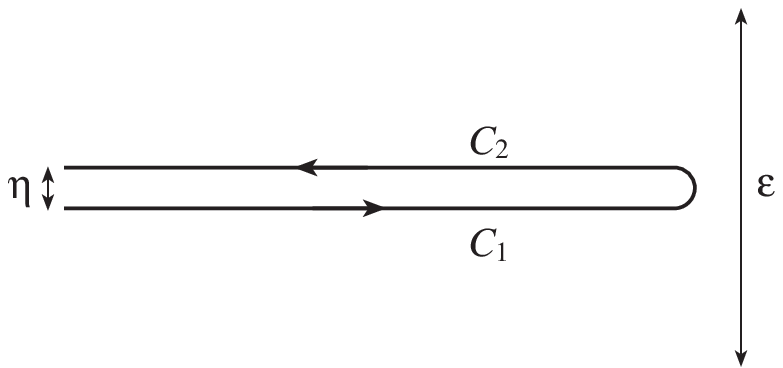}{The zig-zag loop; the loop goes in one direction
along $C_1$ and comes back along $C_2$. The two segments
$C_1$
and $C_2$ are parallel and their distance $\eta$ is less that the
gauge theory UV cutoff $\epsilon$.} If the distance $\eta$ between
$C_1$ and $C_2$ is much less than the UV regularization
$\epsilon$
of the gauge theory, there is one-to-one cancellation between
a Feynman diagram $\Gamma$ which has one of its external leg
ending
on $C_1$ and another diagram $\Gamma'$ which is identical to
$\Gamma$
except that the corresponding leg ends on $C_2$.
Therefore, to all
order in the perturbative expansion, the segments $C_1$
and $C_2$ do not contribute to the expectation value of the Wilson
loop.
On the other hand, if $\dot y^i = |\dot x| \theta^i$
and $\theta^i$ is fixed at a point on $S^5$, a diagram with
a leg coupled to $\dot y^i$ on $C_1$ and one with
the corresponding leg coupled to $\dot y^i$ on $C_2$
add up, rather than cancel each other. The perturbative
computation therefore shows no zig-zag symmetry in this case.

When the coupling $\lambda$ is large, we expect
that $\langle W \rangle$ is related to the minimal surface.
The area functional, and as a matter of fact any other functional
which is an integral over a minimal surface, has zig-zag symmetry.
The proof is simple. If we look at the region $Y \geq \epsilon$,
the minimal surface bounded by a backtracking loop is almost
identical
to the surface bound by the curve without backtracking if the
separation $\eta$
between $C_1$ and $C_2$ is much less than the cutoff
$\epsilon$.
This is illustrated in
\figur{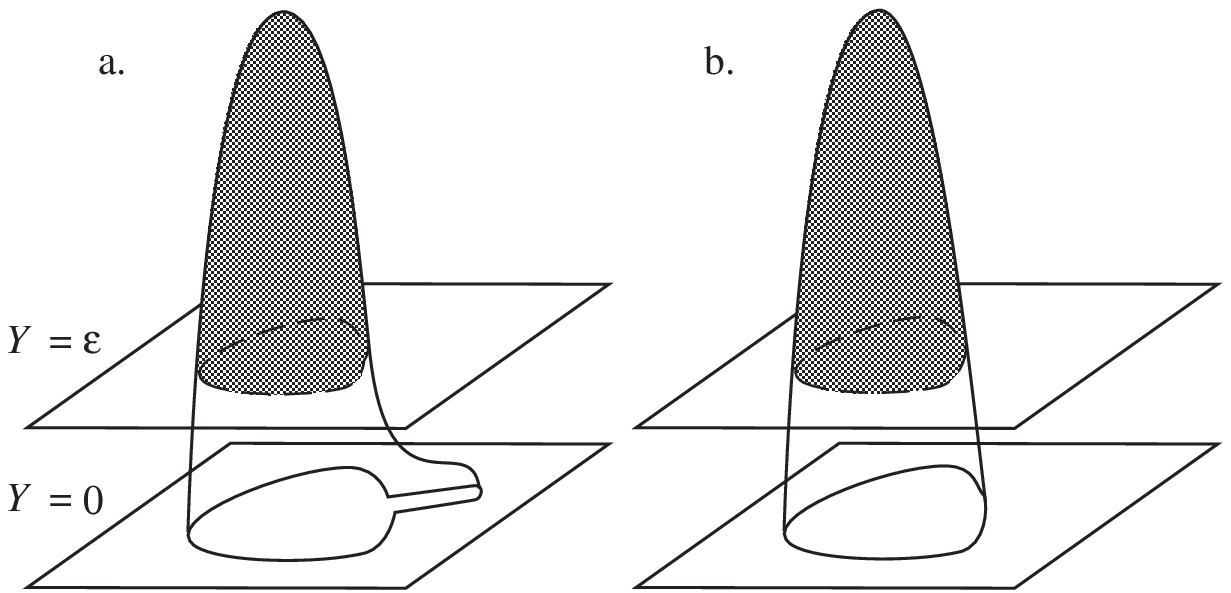}{The area of a loop with a zig-zag (a) is roughly
the same as the loop without it (b).}
Therefore an action on the surface given by an integral over
the part of the surface in $Y \geq \epsilon$ is the same with
or without the backtracking.

At first sight, the zig-zag symmetry of the minimal surface
appears in contradiction with the gauge theory expectation
since we know the minimal surface ending along a smooth loop on
the
boundary of $AdS_5$ obeys the constraint $\dot x^2 = \dot y^2$
and therefore $\dot y^i \neq 0$. In the gauge theory, we
do {\it not} expect zig-zag symmetry when $\dot y^i$
is non-zero and constant. A close examination of the
boundary condition, however, reveals that the situation is more
subtle. It is true that, if we impose the boundary conditions at
$Y=0$, the part of the surface connecting
$C_1$ and $C_2$ do not reach $Y=\epsilon$
 and does not contribute to the regularized area for $Y \geq
\epsilon$.
Therefore zig-zag symmetry holds for $\langle W \rangle$.
This is also the case when we impose the boundary condition
at $Y=\epsilon$. In this case, if $\epsilon \gg \eta$,
the minimal surface goes from
$C_1$ to $C_2$ along the $Y=\epsilon$ surface.
Therefore the contribution of the segments to the regularized area
is proportional to $\eta/\epsilon^2$ times the length of
the segment and vanish in the limit
$\eta \rightarrow 0$.

However the  physical interpretation of the two
computations are quite different. If the boundary conditions
are imposed at $Y=0$, the constraint $\dot x^2 = \dot y^2$
holds provided the segments $C_1$ and $C_2$ are smooth.
On the other hand, if the conditions are imposed on the
$Y=\epsilon$ hypersurface, the minimal surface bounded by $C_1$
and $C_2$
stays within $\eta$ from $Y=\epsilon$,
and $\dot y^2$ vanishes as $\eta/\epsilon \rightarrow 0$.
If we take the latter point of view, the apparent
contradiction with the gauge theory expectation disappears
since the minimal surface in question is related to the
Wilson loop which does not couple to the scalar fields
in the segments $C_1$ and $C_2$. This is exactly the
situation in which zig-zag symmetry arises in the
gauge theory.

One may argue that the boundary condition at $Y=\epsilon$ gives
a more precise definition of the Wilson loop $\langle W \rangle$
as a functional of the loop variables $(x^\mu(s), y^i(s))$.
The Legendre transformation of the area $A$
in section (\ref{corrected-section}), for example, is a way
to define a functional of the momenta $P^i$
evaluated at $Y=\epsilon$ and not at $Y=0$. It does not make
sense to perform this procedure
at $Y=0$ since the factor $1/\epsilon$ in the right-hand side
of (\ref{new-action}) needs to be replaced by $\infty$.
In most of the cases discussed in this paper, whether
we impose the boundary conditions at $Y=0$ or $Y=\epsilon$
does not make much difference since the value of the momenta
$P^i$ stays almost the same in the region $0 \leq Y \leq \epsilon$.
The analysis of  zig-zag symmetry, however, seems to be
an exception to this rule. If we use the boundary condition
at $Y=\epsilon$, the existence of the minimal surface requires
the constraint $\dot y^i(s) = 0$ rather than $\dot x^2 = \dot y^2$
for the backtracking loop, and the result fits well with the
gauge theory expectation. Clearly the regularization dependent
nature
of  zig-zag symmetry needs to be clarified further.

An analysis similar to the one given above leads to the following
observations about the Wilson loop, which we find interesting.
Consider a self-intersecting loop as in \figur{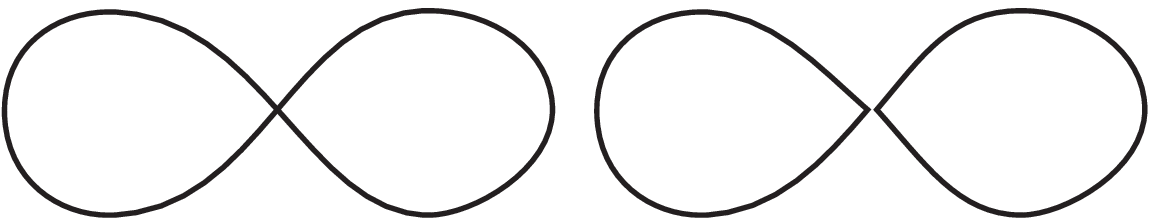}{(a)
A self-intersecting loop which corresponds to a
single trace operator and (b) A pair of loops obtained by
reconnecting the loop at the intersection.}
The area calculated on the minimal surface bound
by the loop (a) is the same as the sum of the two areas bounded
by the separated loops (b).
In the gauge theory, these loops are very different
objects. One is a single trace operator and the other a multi-trace
operator.

We can even connect two distant closed loops by a long neck
without
changing the value of the loop since the minimal surface spanning
the neck region does not contribute to the area.  Graphically this
can
be written as
\beq
\vev{\matrix{\scriptstyle i\cr \scriptstyle j}
\!\!\!=\!=\!=\!=\!=\!=\!=\!\!\!
\matrix{\scriptstyle k\cr \scriptstyle l}}
={1\over N}\delta_{ij}\delta_{kl}
\eeq
This suggests that the parallel transport
$U = {\cal P} \exp (i\int A_\mu dx^\mu )$
along an open curve behaves as
a random matrix. As in the case of the zig-zag symmetry,
if we impose the boundary condition at $Y=\epsilon$, the
minimal surface exists only when $\dot y^i(s)=0$, and
we are considering a loop which does not couple to the
scalar fields in the neck region.

\subsection{Removing the Constraint}
\label{section-lifting}

So far we considered loops of the form (\ref{euclid-phase})
which satisfy the constraint $\dot x^2-\dot y^2=0$.
When the loop has a cusp or an intersection, this constraint
is modified as in (\ref{bad-cusp}) and (\ref{bad-intersection}).
In the gauge
theory, we can define the loop operator for
any $(x^\mu(s), y^i(s))$, not necessarily obeying the
constraint.  Consequently, we need to find a way to calculate
an expectation value of such a loop in $AdS_5$ so that
the relation between the gauge theory and
string theory is complete.

The reason given by Maldacena for the constraint (and also
in Appendix \ref{appendix-phase}) is that the
$W$-bosons are BPS particles and their charges and masses are
related. To break the constraint, one needs a non-BPS object
with an arbitrary mass. Fortunately string theory contains
many such objects. Instead of considering the ground state
of the open string corresponding to the $W$-boson, one may
use excited string states, which have extra
mass from the  string oscillations.
As shown in the Appendix A,
an excited string indeed generates a loop obeying
the modified constraint,
\beq
\dot y^2=\dot x^2 {M^2\over M^2+m^2},
\label{excited}
\eeq
where $M=\epsilon^{-1}$ is the original $W$-boson mass
and $m$ is the mass of the excitations. This makes it possible
to relax the constraint, at least for $\dot x^2 \geq \do y^2$.

For the loop obeying the original constraint $\dot x^2 = \dot y^2$,
the regularized area has the linear divergence of the form
\beq
    A =   \frac{1}{2\pi\epsilon} \oint ds |\dot x| + \cdots
 = \frac{1}{2\pi} \oint M |\dot x| + \cdots.
\eeq
We expect that the corresponding computation using
the string excitation replaces $M$ by
$\sqrt{M^2 + m^2}$ as
\beq
    A = \frac{1}{2\pi}
         \oint ds \sqrt{M^2 + m^2} |\dot x| + \cdots
   = \frac{1}{2\pi \epsilon}
  \oint ds \frac{\dot x^2}{|\dot y|} + \cdots.
\eeq
The Legendre transformation turns this into
\ber
  \tilde{A} &=& A - \frac{1}{2\pi \epsilon}
  \oint ds |\dot y| \nonumber \\
&=&
\frac{1}{2\pi\epsilon}
\oint ds  \left( \frac{\dot x^2}{|\dot y|}
                  - |\dot y| \right)
+ \cdots.
\eer
This shows that the linear divergence is not
completely canceled for $|\dot x| \neq |\dot y|$.
Since a highly excited string state may be sensitive
to stringy corrections, we can trust
this estimate of the linear
divergence only for small deviation
from the constraint. In the following, we will use
an approximate expression for $|\dot x| \sim
|\dot y|$ as
\beq
   \tilde{A}
= \frac{1}{\pi \epsilon}
\oint ds (|\dot x| - |\dot y|) + \cdots.
\label{largeNdiv}
\eeq

\mysection{The Loop Equation}

Since the
expectation value of the Wilson loop is a measure of confinement,
much attention has been  given to calculating them.  In particular, in
the large $N$ limit of gauge theory, they satisfy a closed set of
equations \cite{Makeenko:1979pb}.
In this section, we first give a review of the loop 
equation for pure Yang-Mills theory
(for more details see \cite{Migdal:1984gj,Polyakov:1987ez}).  The
equation is  easy to write down and is
formally satisfied, order by order, in the perturbative
expansion of the gauge theory.
The lattice version of the loop equations are also satisfied in the
non-perturbative
lattice formulation of the theory. However, the only case where 
one can solve explicitly for Wilson loops
is in 2
dimensions. There indeed they do satisfy the loop equation.
We will then formulate the loop equation for the $\cN=4$
super Yang-Mills theory
in 4 dimensions. As far as we know, the loop equation in
this case has not been derived before. We
will find that the BPS condition (\ref{constraint}) will play a crucial
role. We will
discuss details of the construction
in Appendix \ref{appendix-N=4} and present only the general
ideas here.

\subsection{Bosonic Theories}

The action of pure gauge theory in any number of dimensions
is\footnote{The complete action contains a gauge fixing term and
ghosts. Those appear also in the equations of motion, but
can be dropped by a Ward identity \cite{Brandt:1982gz}.}
\beq
\cS={1\over4g_{YM}^2}\int dx\,\Tr F_{\mu\nu}F^{\mu\nu},
\eeq
and the Wilson loop is given by
\beq
W={1\over N}\Tr\,\cP\, \exp\left(i\oint A_\mu dx^\mu\right),
\eeq
where the integral is over a path parametrized by~$x^\mu$.
The main observation is that there is a differential operator on loop
space which brings down the variation of the action $D^\nu
F_{\mu\nu}$
as
\beq
\hat L\,\vev{W}=-i\oint ds\,\dot x^\mu \vev{(D^\nu F_{\mu\nu})^a(s)
\,{1\over N}\Tr\cP\,T^a(s)\exp\left(i\oint A_\mu dx^\mu\right)}
\eeq
where $T^a(s)$ is the generator of the gauge group inserted at 
the point $s$ along the loop.

There are a few equivalent definitions of~$\hat L$. We will use
\beq
\hat L=\lim_{\eta\rightarrow 0}
\oint ds \int_{s-\eta}^{s+\eta} ds'
{\delta^2\over\delta x^\mu(s')\delta x_\mu(s)}.
\label{boson-deriv}
\eeq
As we will explain below, $\eta$ has to be taken much shorter
than the UV cutoff scale $\epsilon$ in order to extract the
term $D^\nu F_{\mu\nu}$.
The insertion of $D^\nu F_{\mu\nu}$ into the loop would
be zero if we use the classical equation of motion,
but quantum corrections produce contact terms.  To see
that, one can write the equations of motion as the functional
derivative of the action $\cS$ and use the Schwinger-Dyson
equations,
$i.e.$ integration by parts in the functional integral,
\ber
\hat L\,\vev{W}
\teq ig_{YM}^2\int\cD A\oint ds\,{1\over N}\Tr\cP\,T^a(s)
\exp\left(i\oint A_\mu dx^\mu\right)
\dot x^\mu(s){\delta e^{-\cS} \over\delta A^{\mu a}(x(s))}
\nonumber\\
\teq -ig_{YM}^2
\vev{\oint ds\,\dot x^\mu(s){\delta\over\delta A^{\mu a}(x(s))}
{1\over N}\Tr\cP\,T^a(s)\exp\left(i\oint A_\mu dx^\mu\right)}.
\phantom{(4.5)}
\eer
The functional derivative $\delta/\delta A_\mu(x(s))$
in this equation is formally evaluated as
\ber
\hat L\,\vev{W}
&=&{\lambda\over N^2}\oint ds\oint ds'\, \delta(x^\mu(s')-x^\mu(s))
\dot x_\mu(s)\dot x^\mu(s') \times \nonumber
\\ &&~~~~~~~~~~~~~\times
\vev{\Tr\,\cP\, T^a(s)T^a(s')
\exp\left(i\oint A_\mu dx^\mu\right)}.
\eer
We then use the relation between the generators of $SU(N)$,
\beq
T^a_{nm}T^a_{kl}
=\delta_{nk}\delta_{ml}-{\delta_{nm}\delta_{kl}\over N}.
\eeq
Ignoring the $1/N$ term, the trace is broken into two.
This gives the
correlation function of two loops. In the large $N$ limit,
the correlator factorizes and we obtain,
\beq
\hat L\,\vev{W}
=\lambda\oint ds\oint ds'\, \delta(x^\mu(s')-x^\mu(s))
\dot x_\mu(s)\dot x^\mu(s') \vev{W_{ss'}}\vev{W_{s's}}.
\label{boson-eqn}
\eeq
Here $W_{ss'}$ is a Wilson-loop that start at~$s$ and goes to~$s'$
and~$W_{s's}$ goes from~$s'$ to~$s$.  They are closed due to
the delta
function\footnote{The delta-function is not sharp,
but is regularized by the cutoff $\epsilon$. That means that the
loops $W_{ss'}$ and $W_{s's}$ are not exactly closed loops, and
the two ends may be separated by a distance $\epsilon$. This does
not contradict  gauge invariance
since one may consider only gauge transformations
which do not vary much over that scale, so the ``almost'' closed
loop are ``almost'' gauge invariant. We expect those loops to be
equal to the closed loops up to $O(\epsilon)$ corrections.}.

Equation (\ref{boson-eqn})
shows that $\hat{L} \langle W \rangle$
receives contributions from self-in\-ter\-sec\-tions of the loop.
Since the derivation of the equation is rather formal,
it is not clear whether we need to count the trivial
case of $s=s'$, in which case $W_{ss'}=1$ and $W_{s's}=W$.
In most of the literature on the loop equation, this trivial
self-intersection is ignored. In any case, 
it can be taken care of by multiplicative renormalization
of the loop operator. In the supersymmetric
gauge theory, the leading contribution
from the trivial self-intersection
cancels when $\dot x^2 = \dot y^2$.

In the definition of the loop derivative $\hat L$, it is important to take
the limit $\eta\rightarrow0$. This procedure isolates
the term $D^\nu F_{\nu\mu}$,
which is a contact term of the double functional derivative. If
$\eta$ is of the order of the UV cutoff $\epsilon$,
there will be other contributions to the
loop equation such as $F_{\mu\nu} F^{\nu\rho} \dot x_\rho$.
When calculating the loop equation in perturbation
theory, we can take $\eta$ to be arbitrarily small,
and in particular $\eta \ll \epsilon$.
This is how we view the loop equation in the continuum theory.
In fact, it was shown that the perturbative expansion of the Wilson
loop solves the loop equation \cite{Brandt:1982gz}.
When we study the loop equation the string in $AdS_5$,
we will consider the same limit $\eta
\rightarrow 0$.

In the lattice regularization, it is not possible to calibrate the
variation of the loop in distance
shorter than the lattice spacing $\epsilon$.
In this case, a different definition
of $\hat L$ is used which does not require taking such a limit.

It is possible to define a loop derivative localized at a point
on the loop, instead of the integrated version considered above.
The entire derivation goes through by simply dropping one $\oint
ds$.

\subsection{Supersymmetric Case}

We briefly summarize how to derive the loop equation in the
supersymmetric theory, leaving the details in Appendix C.
We derive them only for variations from
constrained loops $\dot x^2 = \dot y^2$. One important modification
is due to the extra factor of $i$ in front of the scalars in
the Wilson loop operator in the Euclidean theory,
\beq
W={1\over N}\Tr\,\cP
\exp\left(\oint (iA_\mu \dot x^\mu+\Phi_i\dot y^i)\,ds\right).
\eeq
Another novelty
is the need to include the fermions. The fermions are important
even when the loop equation is evaluated at the body part
$\zeta(s)=0$ of super loop-space
since the fermions appear as source terms
in the equations of motion for the
gauge fields and the scalars. Here we will explain
the effect of the extra $i$. In Appendix
\ref{appendix-N=4}, we will discuss how to deal with the
fermions.

If we define loop derivative
\beq
\hat L=\lim_{\eta\rightarrow 0}
\oint ds \int_{s-\eta}^{s+\eta} ds'
\left({\delta^2\over\delta x^\mu(s')\delta x_\mu(s)}
-{\delta^2\over\delta y^i(s')\delta y_i(s)}\right),
\label{susy-deriv}
\eeq
then the relative minus sign combines with the extra $i$ to give
\ber
\tab\hskip-.25in\hat L\,\vev{W}
\nonumber\\
\teq
-i {g_{YM}^2\over N}\oint ds\,\vev{
\left(\dot x^\mu{\delta\over\delta A^{\mu a}}
-i\dot y^i{\delta\over\delta\Phi^{ia}}\right)
\Tr\cP\,T^a\exp\left(\oint (iA_\mu \dot x^\mu+\Phi_i\dot y^i)\,ds\right)}
\nonumber\\
\teq
\lambda\oint ds\oint ds'
\left(\dot x^\mu(s)\dot x_\mu(s')-\dot y^i(s)\dot y_i(s')\right)
\delta^4(x(s)-x(s'))\vev{W_1}\vev{W_2}.
\phantom{(3.11)}
\label{susy-eqn}
\eer
A simple way to obtain this is by considering the extra $i$ as the
Wick
rotation of the $y^i$ coordinates and repeat the derivation 
from (\ref{boson-deriv}) to (\ref{boson-eqn}).
The right-hand side of the bosonic loop equation contains a cubic
divergence proportional to the circumference of the loop.
In the
supersymmetric case this ``zero-point energy'' cancels
for a smooth loop by the constraint
$\dot x^2 = \dot y^2$.

\subsection{Predictions}

In this subsection, we evaluate the right-hand side of the loop
equation
(\ref{susy-eqn}) for various types of loops. In the next section,
we will compare it with computations of the loop using the minimal
surface spanned by the loop in $AdS_5$.

In the supersymmetric theory, the trivial self-intersection
at $s=s'$ does not contribute to the right-hand side if
the loop is smooth and obeys the constraint $\dot x^2 = \dot y^2$.
This is related to the fact that such a loop
does not require regularization. To be precise, the
constraint only cancels the leading divergence proportional to
$\epsilon^{-3}$. Since
the delta-function in (\ref{susy-eqn}) has a width $\epsilon$,
the Taylor expansion of $x(s')$ at $s'=s$ gives
subleading terms in $\epsilon$ such as
\beq
-{\lambda\over3\epsilon}\oint ds\,(\ddot x^2-\ddot y^2).
\label{ambiguity}
\eeq
However this expression is highly regularization dependent.
Moreover
there are other contributions of the same order due to the fact
that the loops $W_{ss'}$ and $W_{s's}$
are not precisely closed, as explained in the last footnote.
At any rate, these terms are negligible (by a factor $\epsilon$)
compared to the terms we will find at cusps and
intersections, and we will ignore them for the rest of the paper.

For a loop with an intersection, the integral over the
regularized delta-function in the right-hand side of the
loop equation gives
\ber
&&
\lambda (\cos \Omega + \cos \Theta)
\oint ds\oint ds'\, |\dot x(s)||\dot x(s')|
\delta^4_\epsilon\left(x^\mu(s)-x^\mu(s')\right)
\nonumber\\
&=&
\lambda (\cos \Omega + \cos \Theta)
\int_{-\infty}^\infty dx \int_{-\infty}^\infty dx'\,
\delta^4_\epsilon\left(\sin\Omega(x-x')\right) \nonumber \\
&=&\lambda {\cos \Omega + \cos \Theta
\over2\pi\epsilon^2\sin\Omega}.
\eer
It is important to note that the result depends explicitly
on the UV cutoff $\epsilon^{-2}$. Here we have evaluated the
leading term in the $\epsilon^{-1}$ expansion only. There
are subleading terms in the expansion which are comparable
to (\ref{ambiguity}) at the trivial self-intersection.

A cusp also gives an interesting contribution to the loop equation.
This may be regarded as a special case of the trivial
self-intersection. In fact, in the literature, this effect
is ignored together with that of the trivial self-intersection\footnote{
In the lattice formulation, the effect of the cusp to the loop equation
is not seen since there is no local
definition of a cusp.}. In the supersymmetric theory,
the contribution from the trivial self-intersection at a smooth point
on the loop is canceled by the constraint $\dot x^2 = \dot y^2$.
The situation is more interesting at the cusp since
the tangent vector $\dot x^\mu(s)$ is discontinuous there.
If there is a jump on $S^5$, $\dot y^i(s)$ is also discontinuous.
A simple calculation  (identical to (\ref{pert-log}),
where we found the log divergence
in perturbation theory) shows that the cusp contribute to the
right-hand side of the loop equation as
\ber
&&2\lambda (\cos \Omega + \cos \Theta)
\int_{-\infty}^0 dx\int_0^\infty dx'\,
\delta^4_\epsilon\left(\sin\Omega\right(x-x'))\nonumber \\
&=& \lambda {(\pi-\Omega)(\cos \Omega + \cos \Theta)
\over(2\pi\epsilon)^2\sin\Omega}.
\label{predict-cusp}
\eer

To summarize, we can express the loop equation as
\ber
&&\hat L\,\vev{W}
={\lambda\over2\pi\epsilon^2}\left(
\sum_{n:{\rm cusps}}
{(\pi-
\Omega_n)(\cos\Omega_n+\cos\Theta_n)\over2\pi\sin\Omega_n}
\vev{W}+ \right.
\nonumber\\
&&\hskip.2in\left.
+\sum_{m:{\rm intersections}}
{\cos\Omega_m+\cos\Theta_m\over\sin\Omega_m}
\vev{W_m}\vev{\widetilde{W}_m}\right)
+O\left({\lambda\over\epsilon}\right),
\label{predict}
\eer
where $W_m$ and $\widetilde{W}_m$ are Wilson loops
one obtains by detaching the original loop into two
at the intersection point $m$.

\mysection{Loop Equation in $AdS_5 \times S^5$}

\subsection{General Case}

In this section, we will examine whether the computation
of the loop using string theory in $AdS_5$ agrees with the
predictions of the loop equation.
A general form of the loop expectation value is
\beq
\vev{W}=\Delta \exp \left(-\sqrt{\lambda}  \tilde{A}\right).
\label{generalform}
\eeq
We assume that the dependence of the prefactor $\Delta$
on the loop variables is subleading for large $\lambda$.
Since the loop derivative $\hat{L}$ does not commute
with the constraint $\dot x^2 = \dot y^2$, we need an expression
for $\tilde{A}$ when the constraint is not satisfied.
As we saw in section 3.5, 
the exponent $\tilde{A}$ has a linear divergence of the form
\beq
   \tilde{A}(x,y)  = \frac{1}{\pi \epsilon} \oint ds
(|\dot x| - |\dot y|) + \cdots
\label{exponentofw}
\eeq
to the leading order in $(|\dot x| - |\dot y|)$.
The loop derivative is a second order differential operator. When
the
derivatives act on the exponent and bring it down twice,
the result is proportional to
$\lambda$. On the other hand, when they act on $\Delta$ or on the
same
$\tilde{A}$ twice, we get
things only of order $\sqrt\lambda$ or less. In the following,
we will pay attention to the leading term in $\lambda$ only.
The exact expression we have to evaluate is therefore,
\beq
\lambda \lim_{\eta\rightarrow 0}
\oint ds \int_{s-\eta}^{s+\eta} ds'
\left({\delta\tilde A\over\delta x^\mu(s')}
{\delta\tilde A\over\delta x_\mu(s)}
-{\delta\tilde A\over\delta y^i(s')}
{\delta\tilde A\over\delta y_i(s)}\right).
\label{twoderivative}
\eeq
We do not have to include the fermionic derivative. When it acts
once
on a bosonic loop, it gives a fermion whose expectation value is
zero.
There are also non-zero contributions when it acts twice on $\tilde
A$,
but they are subleading in $\lambda$.

Let us evaluate (\ref{twoderivative}). Although the linear
divergence
${1 \over2\pi\epsilon}\oint ds(|\dot x|-|\dot y|)$ in $\tilde{A}(x,y)$
vanishes for the loop obeying the constraint, the variation $\hat{L}$
does not commute with the constraint. Thus the linear
divergence term gives an important
contribution to (\ref{twoderivative}). Since the variation
of the length functional
\beq
L=\oint ds\sqrt{\dot x^2}
\eeq
gives the acceleration $\ddot x^\mu$ (in the parametrization where
$|\dot x|=1$) and the same for $y$, we obtain
\ber
&&\lambda \left( {\delta\tilde A\over\delta x^\mu(s')}
{\delta\tilde A\over\delta x_\mu(s)}
-{\delta\tilde A\over\delta y^i(s')}
{\delta\tilde A\over\delta y_i(s)} \right)\nonumber \\
&=&{\lambda \over\pi^2\epsilon^2}
\left(\ddot x_\mu(s)\ddot x^\mu(s')-\ddot y_i(s)\ddot y^i(s')\right)
+ \cdots.
\label{ads-deriv}
\eer
Note that it has the same divergence, $\epsilon^{-2}$, as
the right-hand side of the loop equation. Moreover the
powers of $\lambda$ match up in the loop equation
and in (\ref{ads-deriv}). The
$\cdots$ in the right-hand side represents variations
of the remaining terms in $\tilde{A}$,
which are finite for a smooth loop.
To compute $\hat{L} \langle W \rangle$,
we integrate (\ref{ads-deriv}) over
$s-\eta \leq s' \leq s+ \eta$.
When the loop is smooth, the acceleration $(\ddot x^\mu, \ddot y^i)$
itself is finite. Therefore, by taking $\eta \rightarrow 0$,
one finds that $\hat{L} \vev{W}=0$ in this case.
This is consistent with the loop
equation. Therefore we reach the first conclusion
that a minimal surface in $AdS_5$ bounded by a
smooth loop solves the loop equation.

\subsection{Loops with Cusps}

If the loop has a cusp of angle $\Omega$, the tangent vector is
discontinuous and $\ddot x$ has a delta-function pointing along
the unit vector bisector $\hat e$
\beq
\ddot x^\mu=2\cos{\Omega\over2}\delta(s)\hat e^\mu.
\eeq
A similar thing happens when $\dot y$ is discontinuous, with the
angle $\Theta$ replacing $\Omega$ in the above.
This delta-function is regularized by $\eta$,
not $\epsilon$, since it is related to the shortest length scale
on which the loop is defined. Thus the integral of (\ref{ads-deriv})
over $s$ and $s'$ gives a non-zero result as
\ber
&& \frac{\lambda}{\pi^2 \epsilon^2} \oint ds\int_{s-\eta}^{s+\eta}ds'
\left(\ddot x_\mu(s)\ddot x^\mu(s')-\ddot y_i(s)\ddot y^i(s')\right)
\nonumber \\
&=&\frac{4\lambda}{\pi^2 \epsilon^2}
\left(\cos^2{\Omega\over2}-\sin^2{\Theta\over2}\right)
\nonumber\\
&=&\frac{2\lambda}{\pi^2 \epsilon^2}
\left(\cos\Omega+\cos\Theta\right)
\label{ads-cusp}
\eer
In comparison with the prediction of
(\ref{predict-cusp}) of the loop equation,
we are missing the factor of $(\pi-\Omega)/\sin\Omega$.
This, however, is not a contradiction. The expression
for the linear divergence term in (\ref{exponentofw})
is an approximation for small $(|\dot x| - |\dot y|)$.
Since $\dot x^2 = (1 + f_0) \dot y^2$ with
$f_0 = f(\Omega/2)$ at the cusp, this approximation
is valid only when $f_0$ is small. Apart from this factor,
(\ref{ads-cusp}) agrees with the prediction of the loop
equation that the cusp gives a non-zero contribution
to the loop equation proportional to $\lambda =
g_{YM}^2 N$ times $\epsilon^{-2}$.

When $(|\dot x| - |\dot y|)$ is not small, the expression
(\ref{exponentofw}) needs to be modified as
\beq
\tilde A(x,y)
= {1 \over\pi\epsilon}
\oint ds|\dot x|G\left({\dot y^2\over\dot x^2}\right)
+ \cdots
\label{modifydivergence}
\eeq
for some function $G(z)$. By repeating the computation that lead to
(\ref{ads-cusp}), we find that the contribution of
the cusp takes the form
\beq
  \hat{L} \exp(-\sqrt{\lambda} \tilde{A})
  =  \lambda
{\cal G}(f_0) (\cos \Omega + \cos \Theta)
  \exp(-\sqrt{\lambda} \tilde{A}) + \cdots,
\label{cuspfinal}
\eeq
where ${\cal G}(f_0)$ is a function related to $G(z)$.
The agreement with (\ref{predict-cusp})
 requires
\beq
  {\cal G}\big(f(\Omega/2)\big)
  = \frac{\pi - \Omega}{8\sin \Omega}.
\eeq
Proving this would be a very strong
evidence for the conjecture.

Loops with cusps have also logarithmic divergences, which could 
contribute to the loop equations. To see that, one may write 
the logarithmically divergent term as
\beq
{1\over2\pi}F(\Omega)\log{L\over\epsilon}
={1\over2\pi}\int ds\int ds' |\dot x(s)||\dot x(s')|
{\sin\varphi\over\pi-\varphi}F(\varphi){1\over(x-x')^2+\epsilon^2}
\eeql{L-zig}
where $\pi-\varphi$ is the angle between $\dot x(s)$ and $\dot x(s')$. 
To check this equation one should integrate over two straight 
lines meeting at a point.
Differentiating (\ref{L-zig}) gives a few terms, among them
\beq
\ddot x(s){1\over\epsilon}{\sin\Omega\over\pi-\Omega}F(\Omega)
\eeql{ads-cusp2}
which has the same divergence as the piece that gave 
(\ref{ads-cusp}).

\subsection{Self-Intersecting Loops}

The situation at a self-intersection is more mysterious
since $\dot x$ and $\dot y$ are both continuous at the
intersection point. However we have problems in our ability to test
the loop equation in this case.
First of all,  $\dot y^i = 0$ at the intersection,
and the function $G(z)$ which appears in the
linear divergence term in (\ref{modifydivergence})
may be singular at $z=|\dot y|/|\dot x| =0$. Since we do not know
about the function $G(z)$ except for its behavior near
$z = 1$, it is difficult to tell whether there is a contribution
from the intersection.

The presence of the unknown
factor $\Delta$ in (\ref{generalform}) makes the
situation worse. As we explained before, the
Wilson loop is
\beq
\vev{W}=\Delta \exp\left(-\sqrt{\lambda}\tilde A\right).
\eeq
For a self-intersecting loop we expect
\beq
\hat L\,\vev{W_{1+2}}=\lambda
{\cos\Omega+\cos\Theta\over\sin\Omega}
\vev{W_1}\vev{W_2},
\eeq
where $W_{1+2}$ is the self-intersecting loop and $W_1$ and
$W_2$ its two pieces. In order for this to be consistent
with the $AdS_5$ computation, we need to find
\beq
\hat L\,\exp\left(-\sqrt{\lambda}\tilde A_{1+2}\right)
=\lambda {\cos\Omega+\cos\Theta\over\sin\Omega}
{\Delta_1\Delta_2\over\Delta_{1+2}}\exp\left(
-\sqrt{\lambda}(\tilde A_1+\tilde A_2)\right).
\eeq
Since we do not know the relation between
the factors $\Delta_1$, $\Delta_2$ and $\Delta_{1+2}$,
a quantitative test is difficult in this case. Though it seems
unlikely that the ration would be zero.

It would be very interesting to determine
the function $G(z)$ what appears in the linear
divergence as it would settle the question as to
whether the intersection gives the contribution
to $\hat{L} \exp(-\sqrt{\lambda}\tilde{A})$
predicted by the  loop equation.

\mysection{Discussion}

The $AdS$/CFT correspondence allows us to calculate 
certain Wilson loops in terms of minimal surfaces in 
anti de-Sitter space. We presented a few reasons why 
only loops satisfying the constraint 
$\dot x^2=\dot y^2$ (generically) are given in terms 
of minimal surfaces. For more general loops we run 
into the problem of inconsistent boundary conditions.

The constrained loops are invariant under half of 
the local 
supersymmetry in super loop-space. As such they are 
BPS objects and are free from 
divergences.
The area of the minimal surface is divergent, so 
it is not the correct functional that yields the Wilson 
loop.  Since the minimal surface satisfies Neumann 
boundary conditions, it's natural to take for the 
action the Legendre 
transform of the area. We showed this yields a
finite result.

In other examples of the $AdS$/CFT correspondence the 
action has to be modified as well. In non-supersymmetric 
cases, such as the near extremal D$3$-brane, the effect of 
adding the boundary term is to subtract $L/(2\pi\epsilon)$. 
The result is finite, but contains a piece proportional to 
the circumference times the radius of the horizon. This may 
be considered a mass renormalization of the $W$-boson. 
The scale of the renormalization is not the UV cutoff,
but rather the scale of supersymmetry breaking. In addition,
if $\dot x^2 \neq \dot y^2$, 
the Wilson loop will contain a linear 
divergence proportional to the UV cutoff.

The surface observables on the $M5$ brane theory, as 
calculated in $AdS_7\times S^4$ have quadratic and 
logarithmic divergences 
\cite{Maldacena:1998im,Berenstein:1998ij,Graham:1999pm}. 
Taking the Legendre transformation will eliminate the 
quadratic divergence, but we are not sure whether it 
will also remove the log divergence.

Recently there were some attempts to go beyond the classical
calculation and include fluctuations of the minimal surfaces
\cite{Forste:1999qn,Greensite:1999jw,Naik}. One
of the goals was to find the ``L\"uscher term,'' the Coulomb
like correction to the linear potential in a confining phase
\cite{Luscher:1980fr}. Any attempt to perform such a 
calculation will require using  the correct Neumann boundary 
conditions on the spherical coordinates, and including the 
appropriate boundary terms.

Finally we formulated the loop equations for those loops, and 
checked if the $AdS$ ansatz satisfies them. For smooth loops, 
due to the supersymmetry, the loop equations should give zero. 
This is indeed the result we find also from the variation of 
the minimal surface.

This calculation actually requires extending the prescription 
to loops that do not satisfy the constraint. We propose that 
the natural extension for small deviation from the constraint 
gives a linear divergence proportional to $\sqrt\lambda L$. 
This term is particularly important when we consider the loop 
equations for loops with cusps. The expected result is finite 
and proportional to $\lambda$. This is in fact what we find, 
but we do not have enough control over the calculation to 
compare the coefficients.

The situation with self-intersecting loops is more 
mysterious, we expect a non-zero answer, but cannot 
reproduce that. There are, however, some reasons why 
this test is more difficult than the other cases. In 
particular, the constraint is broken by a large 
amount at the intersection.

Classical string theory tells us only how to calculate 
loops satisfying the constraint. These are BPS 
objects in loop space, and therefore easier to control. 
As we argued, non-BPS Wilson loops are related to excited 
open strings, but we are unable to evaluate them reliably. 
A similar statement is true for local operators, one 
has control 
only over the chiral operators. Non-chiral operators 
should be given by excited closed string states. 
Despite the large effort devoted to testing the Maldacena 
conjecture, there is still no good understanding of 
non-BPS objects.

\section*{Acknowledgements}
We thank Korkut Bardakci, Sunny Itzhaki, Juan Maldacena,
Joe Polchinski, Bruno Zumino. 
N.D. thanks Theory Group of UC Berkeley and LBNL for
their hospitality.
H.O. thanks the Institute for Theoretical Physics at Santa Barbara,
Department of Mathematics and Physics at University of
Amsterdam,
and Theory Group at CERN, for their hospitality and for providing
excellent working environments.

The work of H.O. was supported in part by the NSF grant 
PHY-95-14797 and the
DOE grant DE-AC03-76SF00098. This work was supported in part
by the NSF under grant No. PHY94-07194.

\appendix

\mysection{Derivation of the Wilson Loop}
\label{appendix-phase}

In this appendix, we will define the coupling of the Wilson
loop to the bosonic fields, $A_\mu$ and $\Phi^i$, in
the ${\cal N}=4$ super Yang-Mills theory. We will pay
special attention to the effect of the Wick rotation
to the Euclidean signature space.
In a gauge theory containing a matter field in the fundamental
representation of the gauge group,
the Wilson loop is derived by writing a correlation
function of the matter fields in terms of the first quantized path
integral over trajectories of the corresponding particle.
The resulting phase factor dictates the proper coupling
of the Wilson loop to the gauge field.
The $\cN=4$ super Yang-Mills theory in 4 dimensions
does not contain such fields. Instead we use $W$-bosons
that appear when we break 
$SU(N+1) \rightarrow SU(N) \times U(1)$.

The bosonic action for the $SU(N+1)$ theory is
\beq
\hat S
=\quart\hat F_{\mu\nu}^2
+\half(\hat D_\mu \hat\Phi_i)^2
-\quart[\hat\Phi_i,\,\hat\Phi_j]^2.
\eeq
By decomposing the gauge group to $SU(N)\times U(1)$ as
\beq
\hat A_\mu = \pmatrix{A_\mu & w_\mu\cr w_\mu^\dagger&a_\mu}
\qquad
\hat \Phi_i = \pmatrix{\Phi_i & w_i\cr w_i^\dagger&M\theta_i},
\eeq
with $\theta^2=1$, the action can be written as
\ber
\hat S
&=&
+\quart F_{\mu\nu}^2
+\half(D_\mu \Phi_i)^2
-\quart[\Phi_i,\,\Phi_j]^2
+\half(\partial_\mu M\theta_i)^2
+(\partial_{[\mu} a_{\nu]})^2
\nonumber\\
&&
+\half w_i^\dagger\left(
(\Phi_k-M\theta_k)^2\delta_{ij}-(\Phi_i-M\theta_i)(\Phi_j-M\theta_j)
\right)w_j
\nonumber\\
&&
+\half((D_\mu-ia_\mu) w_i)^2
+\cdots
\nonumber\\
&=&
S_{SU(N)}
+\half(\partial_\mu M\theta_i)^2
+\quart f_{\mu\nu}^2
+\half((D_\mu-ia_\mu) w_i)^2
\nonumber\\
&&
+\half w_i^\dagger\left(
(\Phi_k-M\theta_k)^2\delta_{ij}-(\Phi_i-M\theta_i)(\Phi_j-M\theta_j)
\right)w_j
+\cdots,
\eer
where $F_{\mu\nu}$ and $f_{\mu\nu}$ are the field strengths of the
$SU(N)$ and $U(1)$ factors respectively. The $\cdots$ in the
action represents terms in higher powers of $w_i$, $etc$.
If $\theta_i$ is in the 1 direction, the mass term
for $w_i$  with $i \neq 1$ becomes
\beq
w_i^\dagger(\Phi_1-M\theta_1)^2w_i
-w_i^\dagger\Phi_i(\Phi_1-M\theta_1)w_1.
\label{pot}
\eeq
with approximate mass eigen-values $\Phi_1-M\theta_1$.
To simplify the following analysis, we replace these
terms with
\beq
w^\dagger(\Phi_1-M\theta_1)^2w.
\eeq

Let us
consider the correlation function
\beq
\vev{w(x)^\dagger w(x) w(y)^\dagger w(y)}.
\eeq
We can integrate over the $w$ field  and find
\ber
&&\hskip-.4in
\int\cD A_\mu \cD \Phi_i \cD w \cD a_\mu
\cD M\theta_i \,
e^{-\hat S}\,
w(x)^\dagger w(x) w(y)^\dagger w(y)
\nonumber\\
&=&
\int\cD M\theta_i \cD a_\mu\,
\exp\left(\int \half(\partial_\mu
M\theta_i)^2+\quart(f_{\mu\nu})^2\right)
\int\cD A_\mu \cD \Phi_i\,
e^{-S_{SU(N)}}\,
\nonumber\\
&&\hskip.5in
\times \bra{x}
{1\over -\half(D_\mu-ia_\mu)^2+\half(\Phi_i-M\theta_i)^2}
\ket{y}
\nonumber\\
&&\hskip1in
\times \bra{y}
{1\over -\half(D_\mu-ia_\mu)^2+\half(\Phi_i-M\theta_i)^2}
\ket{x}.
\eerl{propagator}
The correlation functions in this expression can be
written as
\ber
&&\hskip-.5in
\bra{x}
{1\over -\half(D_\mu-ia_\mu)^2+\half(\Phi_i-M\theta_i)^2}
\ket{y}
\nonumber\\
&=&
\int dT
\bra{x}
e^{T(\half(D_\mu-ia_\mu)^2-\half(\Phi_i-M\theta_i)^2)}
\ket{y}
\nonumber\\
&=&
\int dT
\int_{x(0)=x}^{x(T)=y}\cD x(s)\cD p(s)\,
e^{\int_0^Tds(-i\dot x_\mu p^\mu
-\half(p_\mu+A_\mu+a_\mu)^2-\half(\Phi_i-M\theta_i)^2)}
\nonumber\\
&=&
\int dT
\int_{x(0)=x}^{x(T)=y}\cD x(s)\,
e^{\int_0^Tds(-\half\dot x_\mu^2 + iA_\mu \dot x^\mu
+ia_\mu \dot x^\mu -\half(\Phi_i-M\theta_i)^2)}.
\eerl{manipulate}
Combining everything together and integrating over $y$, we obtain
\ber
&&\hskip-.4in
\int dy\vev{w(x)^\dagger w(x) w(y)^\dagger w(y)}
\nonumber\\
&=&
\int\cD M\theta_i \,
e^{-\int \half(\partial_\mu M\theta_i)^2}
\int dT
\int_{x(0)=x}^{x(T)=x}\cD x(s)\,
e^{-\half\int_0^Tds(x_\mu^2 + M^2)}
\nonumber\\
&&
\int \cD a_\mu\,
e^{\int\quart(f_{\mu\nu})^2}
e^{\oint ds\,ia_\mu \dot x^\mu}
\int\cD A_\mu \cD \Phi_i\,
e^{-S_{SU(N)}}\,
e^{\int ds(iA_\mu \dot x^\mu-\half\Phi_i^2+M\Phi_i\theta^i)}.
\nonumber\\
\eerl{one-theta}

Lets examine (\ref{one-theta}) carefully.  The first term $M^2\int
\half(\partial_\mu \theta_i)^2$ is the action of the $\theta_i$ field,
which for large $M$ becomes classical.
The second term includes an integral over all the closed paths
through
$x$.  To define the Wilson loop we just look at one such path,
leaving
the integration over paths for latter.  The next term in the exponent
breaks reparametrization invariance and will set $\dot
x_\mu^2=\theta_i^2$, as shown below.
The next term is the action for the Abelian gauge field on the single
brane and the effect of the Wilson loop on it. Since $N\gg1$ and we
are taking the probe approximation, we should ignore this term.
As we'll see, for large $M$ the $\Phi^2$ term will be subleading, so
the last term is simply the Wilson loop
\beq
\langle W(x^\mu,\theta_i) \rangle
=
\int\cD A_\mu \cD \Phi_i\,
e^{-S_{SU(N)}}\,
e^{\int ds(iA_\mu \dot x^\mu - \Phi_i\theta^i)}.
\eeq
The term with $\dot x^2+M^2$ is not reparametrization invariant.
When we perform the integral over different parametrizations of the
same path (including the integral over $T$) we find a saddle point.
A general parametrization is $s \rightarrow \tilde s(s)$
such that $\tilde s(0)=0$ and $\tilde s(\tilde T)=T$.
To integrate over different parametrizations,
we can perform the path integral over
$c(s)=d\tilde s/ ds$ with action
\beq
-\int_0^Tds\,
\half\left({1\over c}\dot{\tilde x}_\mu^2
+cM^2\right)
+\int_0^Tds
(iA_\mu \dot {\tilde x}^\mu
-c\half\Phi_i^2
+cM\Phi_i\theta^i).
\eeq
For large $M$ the first term dominates, so it will pick the
saddle point
\beq
c(s)^2={\dot x_\mu^2\over M^2},
\eeq
and indeed the $\Phi^2$ piece in the loop drops out.

Combining them together, we obtain,
\ber
&&\hskip-.4in
\int dy\vev{w(x)^\dagger w(x) w(y)^\dagger w(y)}
\nonumber\\
&=&
\int\tilde\cD x(s)\,
e^{-\int ds\,M|\dot x|}
\int\cD A_\mu \cD \Phi_i\,
e^{-S_{SU(N)}}\,
e^{\int_0^1ds
\left(iA_\mu \dot x^\mu
+|\dot x|\Phi_i\theta^i\right)}.
\eer
The integral $\int ds|\dot x|M$ is the length of the loop
times the mass $LM$. Since it is a $c$-number independent
of $\lambda$, we can ignore it as subleading in the
large $\lambda$ analysis in this paper. For the same
reason,  possible determinant factors are also
neglected in the above.

The calculation above can also be done in Lorentzian signature.
The
difference is an extra $i$ in (\ref{manipulate})
\ber
&&\hskip-.5in
\bra{x}
{1\over -\half(D_\mu-ia_\mu)^2+\half(\Phi_i-M\theta_i)^2}
\ket{y}
\nonumber\\
&=&
\int dT
\bra{x}
e^{iT(+\half(D_\mu-ia_\mu)^2-\half(\Phi_i-M\theta_i)^2)}
\ket{y}.
\eer
The rest of the calculation carries through with this $i$ showing
up in different places. The final result is
\ber
&&
\int\cD \theta_i \,
e^{i\int \half(\partial_\mu \theta_i)^2}
\int\tilde\cD x(s)\,
e^{i\int_0^1ds\,M|\dot x|}
\int \cD a_\mu\,
e^{i\int\quart(f_{\mu\nu})^2}
e^{i\int ds\,a_\mu \dot x^\mu}
\nonumber\\
&&\hskip.5in
\int\cD A_\mu \cD \Phi_i\,
e^{iS_{SU(N)}}\,
e^{i\int_0^1ds
\left(A_\mu \dot x^\mu
+|\dot x|\Phi_i\theta^i\right)}
\eer
though it is less clear now why the term $i(\dot x^2+M^2)$ should
dominate the path integral to set the saddle point.

Instead of the $W$-boson,
we may consider a more general particle with
an arbitrary mass with a propagator
\ber
{1\over \half(D_\mu-ia_\mu)^2
+\half(\Phi_i-M\theta_i)^2
+\half m^2}.
\eer
By the same calculation as above, we obtain the exponent
\beq
-\int_0^1ds
\sqrt{\dot x_\mu^2(M^2+m^2)}
+\int_0^1ds\left(
iA_\mu \dot x^\mu
+{M|\dot x|\over\sqrt{M^2+m^2}}\Phi_i\theta^i\right).
\eeq
Excited states of the open strings have this propagator and can
be used to construct loops with $\dot x^2\neq \dot y^2$.

So far $\theta$ is a constant. To construct
loops which move in the $\theta$
directions, we have to use many probe D-branes, one for each
value
of $\theta$ the loop goes through. We start with $SU(N+M)$ and
break
to $SU(N)\times SU(M)$ which will then be broken to
$SU(N)\times U(1)^M$. Likewise one should be able to  couple the
loop to
the fermions to get the supersymmetric loops used in Appendix
\ref{appendix-N=4}

\mysection{Area of a Cusp}
\label{appendix-cusp}

\subsection{At One Point on $S^5$}

Here we study the minimal surface near a cusp. We consider
a loop on a 2-dimensional plane in 4 dimensions,
staying at the same point on $S^5$. We take the
opening angle of the cusp to be $\Omega$.
We choose radial coordinates $r$ and $\varphi$ on the plane
and use them to parametrize the worldsheet also.
The boundary conditions are (using
the first regularization discussed in \ref{section-area})
\beq
Y(r,0)=Y(r,\Omega)=0
\eeq
To study the behavior of the surface near the cusp, we can use
scale
invariance to set
\beq
Y(r,\varphi)={r\over f(\varphi)}
\eeq
Using this ansatz, the area is
\beq
A={1\over2\pi}\int dr\,d\varphi{1\over r}
\sqrt{f^4+f^2+f^{\prime2}}
\eeq
This reduces the minimal surface to a one-dimensional
problem with the effective Lagrangian
\beq
L=\int d\varphi\sqrt{f^4+f^2+f^{\prime2}}.
\eeq
Since $L$ does not depend explicitly on $\varphi$,
the energy $E$ given by
\beq
E={f^2+f^4\over\sqrt{f^4+f^2+f^{\prime2}}}
\label{energy}
\eeq
is conserved.
At the minimum of $f$, the energy is given by
\beq
E=f_0\sqrt{1+f_0^2},\qquad (f_0 = f(\Omega/2)).
\eeq
Substituting this back in (\ref{energy}),
\ber
{\Omega\over2}
&=&\int_0^{\Omega/2} d\varphi \nonumber\\
&=&f_0\sqrt{1+f_0^2}\int_{f_0}^\infty
{df\over f\sqrt{(1+f^2)(f^2-f_0^2)(f^2+f_0^2+1)}}\nonumber \\
&=&f_0\sqrt{1+f_0^2}\int_0^\infty
{dz\over (z^2+f_0^2)\sqrt{(z^2+f_0^2+1)(z^2+2f_0^2+1)}}
\nonumber\\
&=&{i\over f_0}\Pi\left(\arcsin i\infty, {\sqrt{1+2f_0^2}\over f_0},
\sqrt{1+2f_0^2\over1+f_0^2}\right),
\label{pielliptic}
\eer
where $\Pi$ is an elliptic integral of the third kind.
The regularized action is then
\ber
L&=&\int_{r \geq \epsilon f(\varphi)}
 d\varphi\sqrt{f^4+f^2+f^{\prime2}} \nonumber \\
&=&\int dz\sqrt{z^2+f_0^2+1\over z^2+2f_0^2+1}
\nonumber\\
&=&i\sqrt{1+f_0^2}E\left(
\arcsin i\sqrt{{r^2\over\epsilon^2}-f_0^2\over 1+2f_0^2},
\sqrt{1+2f_0^2\over 1+f_0^2}\right)
\eer
where $E$ is an elliptic integral of the second kind. For small
$\epsilon$, it diverges linearly as
$2 r/\epsilon -F(\Omega)$. The function $F$ is obtained
by solving (\ref{pielliptic}) for $f_0$ as a function of $\Omega$
and substituting it into $L$ in the above.
The total area is
\beq
A={1\over2\pi}
\int^L dr {1\over r}\left({2r\over\epsilon}-F(\Omega)\right)
=
{2L\over2\pi \epsilon}-{1 \over 2 \pi} F(\Omega)\log{L\over\epsilon}.
\eeq
This is the regular linear divergence plus a logarithmic divergence.
After the Legendre transformation, we obtain
\beq
\tilde A
=-{1\over2\pi}F(\Omega)\log{L\over\epsilon}.
\eeq

\subsection{With a Jump on $S^5$}

The same analysis can be done for a loop which jumps,
at the cusp, to a different point on $S^5$
with a relative angle $\Theta$.
We parametrize the string worldsheet
by $r$ and $\theta$, where $\theta$
is a coordinate along the large circle connecting the 2
different points on $S^5$.
Because of scale invariance, we can set
\beq
   Y(r,\theta) = \frac{r}{\tilde{f}(\theta)},
\eeq
for some function $\tilde{f}(\theta)$.
The other angular parameter $\varphi$ is
a function of $\theta$ only. The area is therefore
\beq
A=\int dr\,d\theta{1\over r}
\sqrt{\tilde f^{\prime2}+(1+\tilde{f}^2)(1+\tilde{f}^2\varphi^{\prime2})}.
\eeq
The problem is integrable since there
 are two conserved quantities,
\beq
E={1+\tilde{f}^2\over
\sqrt{\tilde{f}^{\prime2}+(1+\tilde{f}^2)(1+\tilde{f}^2\varphi^{\prime2})}}
\qquad
J={(1+\tilde{f}^2)\tilde{f}^2\varphi'\over
\sqrt{\tilde{f}^{\prime2}+(1+\tilde{f}^2)(1+\tilde{f}^2\varphi^{\prime2})}}
.
\eeq
In general the result cannot be written
in terms of elliptic integrals, and we will leave it to the over
motivated reader to find simple expressions for those integrals.
If we set $\Omega=\pi$, there is no cusp in the $x$ plane.
In this case, the integrals are simplified, and
the results are expressed in terms of the elliptic integrals.

\mysection{Details of Loop Equation in $\cN=4$ Super Yang-Mills
Theory}
\label{appendix-N=4}

The bosonic part of the Euclidean Wilson loop is
\beq
W={1\over N}\Tr\,\cP
e^{\int (iA_\mu \dot x^\mu+\Phi_i\dot y^i)\,ds}
\eeq
We can define the bosonic part of the loop derivative to be
\beq
\hat L=\lim_{\eta\rightarrow 0}
\int ds \int_{s-\eta}^{s+\eta} ds'
\left({\delta^2\over\delta x^\mu(s')\delta x_\mu(s)}
-{\delta^2\over\delta y^i(s')\delta y_i(s)}\right)
\eeq
The extra $i$ in front of $\Phi_i\dot y^i$ in the exponent conspires
with the relative minus sign in the loop derivative to give the
bosonic part of the equations of motion
\ber
\hat L\, \vev{W}&=&-i\int ds\,\left<
\left(\dot x^\mu(D^\nu F_{\mu\nu})^a
+i\dot x^\mu[\Phi^i,D_\mu\Phi_i]^a
+i\dot y^i(D^\nu D_\nu\Phi_i)^a
\right.\right.
\nonumber\\
&&\hskip.5in
\left.\left.
-i\dot y^i[\Phi^j,[\Phi_i,\Phi_j]]^a\right)
\Tr\cP\,T^a(s)e^{\int (iA_\mu \dot x^\mu+\Phi_i\dot y^i)\,ds}\right>
\eer
This is a linear combination of the bosonic equations of motion for
$A_\mu$ and $\Phi^i$, but we are missing source terms due
to the fermions. What we would like to do here is to modify
the functional differential operator $\hat{L}$, including
derivatives of fermionic variables, so that the full equations
of motion are reproduced. With such $\hat{L}$, the loop
equation can be written as
\ber
\hat L\,\vev{W}&=&-i {g_{YM}^2\over N}\int ds\,\vev{
\left(\dot x^\mu{\delta\over\delta A^{\mu a}}
-i\dot y^i{\delta\over\delta\Phi^{ia}}\right)
\Tr\cP\,T^a(s)e^{\int (iA_\mu \dot x^\mu+\Phi_i\dot y^i)\,ds}}
\nonumber\\
&=&\lambda\int ds\int ds'
\left(\dot x^\mu(s)\dot x_\mu(s')-\dot y^i(s)\dot y_i(s')\right)
\delta^4(x(s)-x(s'))W_1W_2
\nonumber\\
\eer

The Euclidean super Yang-Mills theory has
fermionic fields $\Psi$ which are Euclidean
Majorana fermions
\cite{Nicolai:1978vc} with 16 complex components.
The gamma matrices $\Gamma_M$ satisfy the Dirac algebra in 10
dimensions with signature (10,0),
with the index $M=(\mu,i)$.
The loop is parametrized by $(x^\mu(s),~y^i(s))$ and their
superpartner $\zeta(s)$ coupling to the gauginos $\Psi$.

A natural choice for the supersymmetrized loop is
\beq
W={1\over N}\Tr\,\cP\left[
e^{\int\bar\zeta(s)Q\,ds}
e^{\int (iA_\mu \dot x^\mu+\Phi_i\dot y^i)\,ds}
e^{-\int\bar\zeta(s)Q\,ds}
\right]
\eeq
Here $Q$ is the generator of supersymmetry of the gauge theory,
which acts as
\ber
[Q,A_M]&=&{i\over2}\Gamma_M\Psi
\nonumber\\
\{Q,\Psi\}&=&-\quart\Gamma_{MN}F^{MN},
\eer
where we have combined the gauge field $A_\mu$ and the
scalars $\Phi^i$ into the 10-dimensional gauge field
$A_M$ and computed the field strength $F_{NM}$.
One may also include
\beq
[Q,\dot x_M]={i\over4}\Gamma_M\dot\zeta
\eeq
in the exponent,
but it does not affect our analysis since we will only be interested at
the top component of the Grassmann algebra and at the end of the
calculation we set $\zeta=0$.
The exponent of the Wilson loop is therefore given by
\ber
&&
e^{\bar\zeta Q}(iA_\mu \dot x^\mu+\Phi_i\dot y^i) e^{-\bar\zeta Q}
\nonumber\\
&=&
(iA_\mu \dot x^\mu+\Phi_i\dot y^i)
-\half\bar\zeta(\dot x^\mu\Gamma_\mu-i\dot
y^i\Gamma_i)\Psi\nonumber\\
&&-{1\over 16}
\dot x^\mu F^{\nu\rho}\,\bar\zeta\Gamma_\mu\Gamma_{\nu\rho}\zeta
+\ldots.
\eerl{superphase}

We will write the loop equation only for loops satisfying the
constraint $\dot x^2 = \dot y^2$. Therefore
$\dot x^M\Gamma_M=\dot x^\mu\Gamma_\mu-i\dot y^i\Gamma_i$ is
nilpotent.  In this case, it is useful to work in the basis where
\ber
\dot x^\mu\Gamma_\mu-i\dot y^i\Gamma_i=\pmatrix{0&0\cr2|\dot
x|&0}
\qquad
\zeta={1\over\sqrt{|\dot x|}}\pmatrix{\zeta_1\cr\zeta_2}
\qquad
\Psi=\pmatrix{\psi_1\cr\psi_2}
\eer
and
\beq
\bar\zeta=\zeta^T\cC
\eeq
where $\cC$ is the charge conjugation matrix. The Majorana
spinor in Lorentzian signature space satisfies the reality condition
$\bar\zeta=\zeta^\dagger\Gamma^0$.  In the Euclidean case,
we do not impose any reality condition \cite{Nicolai:1978vc}.
The exponent of the loop (\ref{superphase}) in this basis
becomes
\ber
(iA_\mu \dot x^\mu+\Phi_i\dot y^i)
-\sqrt{|\dot x|}\,\bar\zeta_1\psi_1
+{1\over 8}\sqrt{|\dot x|}
F^{NM}\,\bar\zeta_1\Gamma_{NM}\zeta
+\ldots.
\eer
By applying the fermionic derivative operator
\beq
{1\over|\dot x|}
{\delta\over\delta\zeta(s')}{\delta\over\delta\bar\zeta(s)}W
\sim
{\delta\over\delta\zeta_1(s')}{\delta\over\delta\bar\zeta_1(s)}W,
\eeq
we obtain the desired combination for the source terms
in the equation of motion,
\beq
|\dot x|\bar\psi_1\psi_1
=
\bar\Psi(\dot x^\mu\Gamma_\mu+i\dot y^i\Gamma_i)\Psi.
\eeq
All other terms contain at least one $\zeta(s)$ and is
not relevant for our analysis of the loop at $\zeta=0$.
Thus we found the supersymmetric loop derivative
defined by
\beq
\hat L=\lim_{\eta\rightarrow 0}
\int ds \int_{s-\eta}^{s+\eta} ds'
\left({\delta^2\over\delta x^\mu(s')\delta x_\mu(s)}
-{\delta^2\over\delta y^i(s')\delta y_i(s)}
+{\delta\over\delta\zeta(s')}{\delta\over\delta\bar\zeta(s)}\right)
\eeq
produces the variation of the action. For
the loop at $\zeta=0$, this completes
the loop equation for the ${\cal N}=4$ super Yang-Mills theory.

\end{document}